\documentclass[twocolumn]{aastex63}

\usepackage{CJK}
\usepackage{textcomp}
\usepackage{makecell}
\usepackage{tabularx}
\usepackage{threeparttable}
\usepackage{multirow}
\usepackage{rotating}
\usepackage{lineno}

\usepackage{supertabular}
\usepackage{longtable}
\usepackage{amsmath}

\received{xxx}
\revised{xxx}
\accepted{xxx}

\newcommand{\ttilde}{\raisebox{0.5ex}{\texttildelow}}

\shorttitle{FTS exoplanet imaging}
\shortauthors{Zhang, Bottom, and Serabyn 2023}

\graphicspath{{./}{figures/}}

\begin{document}
\begin{CJK*}{UTF8}{gbsn}

\title{Direct detection and characterization of exoplanets using imaging Fourier transform spectroscopy}

\correspondingauthor{Jingwen Zhang}
\email{jingwen7@hawaii.edu}

\affiliation{Institute for Astronomy, University of Hawaii at Manoa\\
2680 Woodlawn Dr. \\
Honolulu, HI 96822, USA}

\author[0000-0002-2696-2406]{Jingwen Zhang(张婧雯)}
\altaffiliation{NASA FINESST Fellow}
\affiliation{Institute for Astronomy, University of Hawaii at Manoa\\
2680 Woodlawn Dr. \\
Honolulu, HI 96822, USA}

\author{Michael Bottom}
\affiliation{Institute for Astronomy, University of Hawaii at Manoa\\
640 N. Aohoku Pl,\\ Hilo, HI 96720, USA}

\author{Eugene Serabyn}
\affiliation{Jet Propulsion Lab, California Institute of Technology\\
4800 Oak Grove Dr, Pasadena, CA 91109, USA}
\begin{abstract}

 Space-based direct imaging provides prospects for detection and spectral characterization of exoplanets at optical and near-infrared wavelengths. Integral field spectrographs (IFS) have been historically baselined for these mission concepts. However, multiple studies have revealed that detector noise is a serious obstacle for such instruments when observing extremely faint targets such as Earth-like planets.  Imaging Fourier transform spectrographs  (iFTS) are generally less sensitive to detector noise, and have several other compelling features such as simultaneous imaging and spectroscopy, smaller-format detector requirements, and variable spectral resolving power.  To date, they have not been studied as options for such missions. 


Here, we compare the capabilities of integral field spectrographs and imaging Fourier transform spectrographs for directly obtaining spectra from an Earth-like planet using both analytic and numerical models. Specifically, we compare the required exposure times to achieve the same signal-to-noise ratio with  the two architectures over a range of detector and optical system parameters.  We find that for a 6-meter telescope, an IFS outperforms an iFTS at optical wavelengths due to the effects of distributed photon noise. In the near-IR, the relative efficiency of an IFS and iFTS depends critically on the instrument design and detector noise. An iFTS will be more efficient than an IFS if the readout noise of the near-IR detector is above \ttilde $2-3 \ e^{-}\ pix^{-1}\ \rm{frame}^{-1} (t_{\rm{frm}}=1000s)$, which correspond to half to one-third of the state-of-art detector noise. However, if the readout noise is reduced below this threshold, the performance of an IFS will experience a substantial improvement and become more efficient.  These results motivate consideration of an iFTS as an alternative option for future direct imaging space missions in the near-IR.

\end{abstract}


\section{Introduction}\label{sec:intro}

Spectroscopic observations of exoplanet atmospheres are a powerful tool to measure chemical compositions and abundances, and provide key insights into the formation, evolutionary history and habitability of exoplanets. In the fortuitous case of an exoplanet transit, transmission spectroscopy provides one avenue for measuring the absorption features imprinted by the planetary atmosphere on the stellar spectrum \citep{Charbonneau2002}. Direct imaging offers another way to obtain the reflection or thermal emission spectrum of an exoplanet by suppressing the host star's light using a coronagraph or a starshade. Although simple in principle, direct imaging is challenging due to the large flux ratio and small separation between a planet and its host star. For example, a Jupiter twin orbiting a Sun-like star at 10 pc has a planet-star flux ratio on the order of $10^{\text{-}9}$ in optical at a separation of $0.52^{\arcsec}$. Using large ground-based telescopes, thermal emission spectra of several young gas-giant exoplanets at wide orbits  have been obtained, for which planet-star flux ratios are on the order of $10^{\text{-}4}$ to $10^{\text{-}6}$. Such systems include beta Pic b \citep{lagrange2009probable}, 51 Eri b \citep{Macintosh2015} and the four planets around HR 8799 \citep{Marois2008,Barman2011,Konopacky2013}. 

Space-based direct imaging telescopes are expected to bring a breakthrough in sensitivity by avoiding the limitations imposed by the Earth's atmosphere. An important development in this direction is the coronagraph instrument on the \textit{Nancy Grace Roman Space Telescope} (\textit{Roman-CGI}), which will achieve optical contrast ratios on the order of $10^{\text{-}9}$ \citep{Romancgiparams2022}.\textit{Roman-CGI} is expected to detect and characterize gas giant exoplanets around nearby stars. NASA has also studied mission concepts such as the Habitable Exoplanet Observatory (HabEx) \citep{Gaudi2018} and Large UV Optical Infrared Surveyor (LUVOIR) \citep{LUVOIRTeam}. The Decadal Survey on Astronomy and Astrophysics 2020 (Astro 2020, \citealt{Decadal2021}) recommended a decade long effort to mature technologies for space-based high contrast imaging, specifically towards a 6 meter high-contrast direct imaging telescope, currently referred to as the Habitable Worlds Observatory (HWO), with the goal of detecting and characterizing $\sim 25$ terrestrial planets around nearby sun-like stars. 

Spectroscopy at ultraviolet (UV), visible and near-infrared (NIR) wavelengths is key to atmospheric characterization of these planets, especially to measure abundances of biosignature gases such as $\rm{O_{2}}$, $\rm{O_{3}}$, $\rm{CH_{4}}$.  Integral field spectrographs (IFS) were baselined for HabEx and LUVOIR, and were the choice for \textit{Roman-CGI} until  descoped \citep{stark2014,LUVOIRTeam,gong2019,Martin2019, Gaudi2020}. By placing an lenslet array in the focal plane, an IFS can transfer the light at each individual location to prisms or gratings, which disperse the target photons into different detector pixels to retrieve the spectral content \citep{Courtes1982}. Therefore, one can use an IFS to obtain spatially resolved spectra from a two dimensional image. However, IFS performance is severely compromised by detector noise when observing faint targets like terrestrial exoplanets. \cite{Robinson2016} and \cite{Lacy2019} presented detailed simulations of space-based coronagraphs equipped with integral field spectrographs, and both studies concluded that detector noise dominates the overall error budget, followed by photon noise of the target, residual starlight, and zodiacal light. For example, for an Earth twin at 5 pc observed with a 2 m telescope, \cite{Robinson2016} compute that the planet signal rate of a spectral element at 546-554 nm is $\ttilde 2\ \rm{e^{\text{-}}\ hr^{\text{-}1}}$, whereas the dark current is $\ttilde240\ \rm{e^{\text{-}}\ hr^{\text{-}1}}$, requiring roughly 5000 hrs to achieve a SNR of 5. Similarly, \cite{Lacy2019} find that for the \textit{Roman-CGI}, the dark current is between 1.5 and 3 orders of magnitude higher than the planet photon rate from a Jupiter twin at 10 pc for all wavelengths between 550 and 1000 nm. This sensitivity to detector noise is exacerbated by an IFS dispersing photons into a relatively large number of pixels per spectral element.

An imaging Fourier transform spectrograph (iFTS) offers a distinct method for acquiring spatially resolved spectra. Unlike an integral field spectrograph, which disperses light spatially, an iFTS modulates light intensity over time using a movable mirror to create interference patterns. This approach doesn't disperse photons spatially, resulting in higher photon noise and reduced performance for bright targets under photon-limited conditions. However, the absence of dispersion means an iFTS needs fewer pixels and is less affected by detector noise in scenarios where the detector's limitations are the main concern. Additionally, an iFTS can use a smaller detector size that is determined by the desired field of view rather than the spectral resolving power and bandwidth.  This offers flexibility in spectral resolving power, ranging from a broad single channel (eg, like an imaging camera) to very high in the case of a long mirror scan.

Despite its advantages, the iFTS brings increased complexity in data acquisition, analysis, and instrument construction, including the mechanism for translating the mirror. Despite these challenges, the fundamental advantages in sensitivity in the detector noise limited regime make an iFTS an intriguing choice for direct imaging and characterization of faint exoplanets.  Many of these challenges have been solved in previous iFTS instruments, which have a long heritage in space and ground astronomy.


First introduced by Peter Fellgett in 1949, Fourier transform spectrographs have been successfully applied to the field of planetary exploration on the Mariner, Voyager, and Cassini spacecrafts \citep{Hanel1972, Hanel1979, Flasar2004} and on ground-based telescopes \citep{Weisstein1996}.  They have also found use in cosmology, such as in the FIRAS instrument of the Cosmic Background Explorer satellite \citep{Mather1990}. An imaging Fourier transform spectrograph was also considered for the James Webb Space Telescope \citep{Graham1998}, but was not selected for the final instrument suite as improvements in detector performance, particularly the advent of the HAWAII arrays \citep{hodapp1996}, obviated the advantages of the iFTS for the typical targets observed by Webb. Successful ground-based implementations can be found in the solar Fourier Transform Spectrometer on Kitt Peak National Observatory \citep{brault1978solar} and the SITELLE instrument on the Canada-France-Hawaii Telescope (CFHT, \citealt{Grandmont2012}).   Unlike IFS's which routinely obtain exoplanet spectra on ground based telescopes, IFTS's have not yet been demonstrated to work with high contrast imaging systems, though there is at least one pathfinder instrument, SPIDERS, planned for deployment at Subaru telescope in the summer of 2024 \citep{maroisinprep, thompson2023lights}.  However, there is no study to date of using an iFTS for space-based exoplanet spectroscopy and comparing its performance with that of an IFS.

 In this paper, we compare the performance of an iFTS to an IFS in the same spirit as the work of \cite{Robinson2016} and \cite{Lacy2019}, using analytic models and detailed numerical simulations over a range of telescope and instrument parameters. Specifically, we use the required exposure times to achieve a signal-to-noise ratio of 5 on the continuum as a metric to compare the capabilities of an iFTS and an IFS.
 
 The outline of the paper is as follows: in Section~\ref{sec:FTSconcept}, we provide a brief review of the iFTS concept. Section~\ref{sec:me} describes the analytical model and numerical simulations, while Appendices present the detailed derivation and calculation of our models.  Section \ref{sec:instru} outline the instrument parameters used in our simulations. Section~\ref{sec:resu} presents our results of comparing the performance of an iFTS and an IFS. Section~\ref{sec:diss} discusses further avenues for research and Section~\ref{sec:concs} summarizes our conclusions.

\section{iFTS Concept} \label{sec:FTSconcept}

Most spectrographs are hardware implementations of the Wiener-Khinchin theorem, which states that the Fourier transform of the autocorrelation of a signal yields its power spectral density, ie, its spectrum \citep{kauppinen2001fourier}.  In a grating spectrograph, for example, the autocorrelation is computed \textit{spatially} through the displacement of the signal by the successive grooves of the grating.

\begin{figure}
    \centering
    \includegraphics[width=\linewidth]{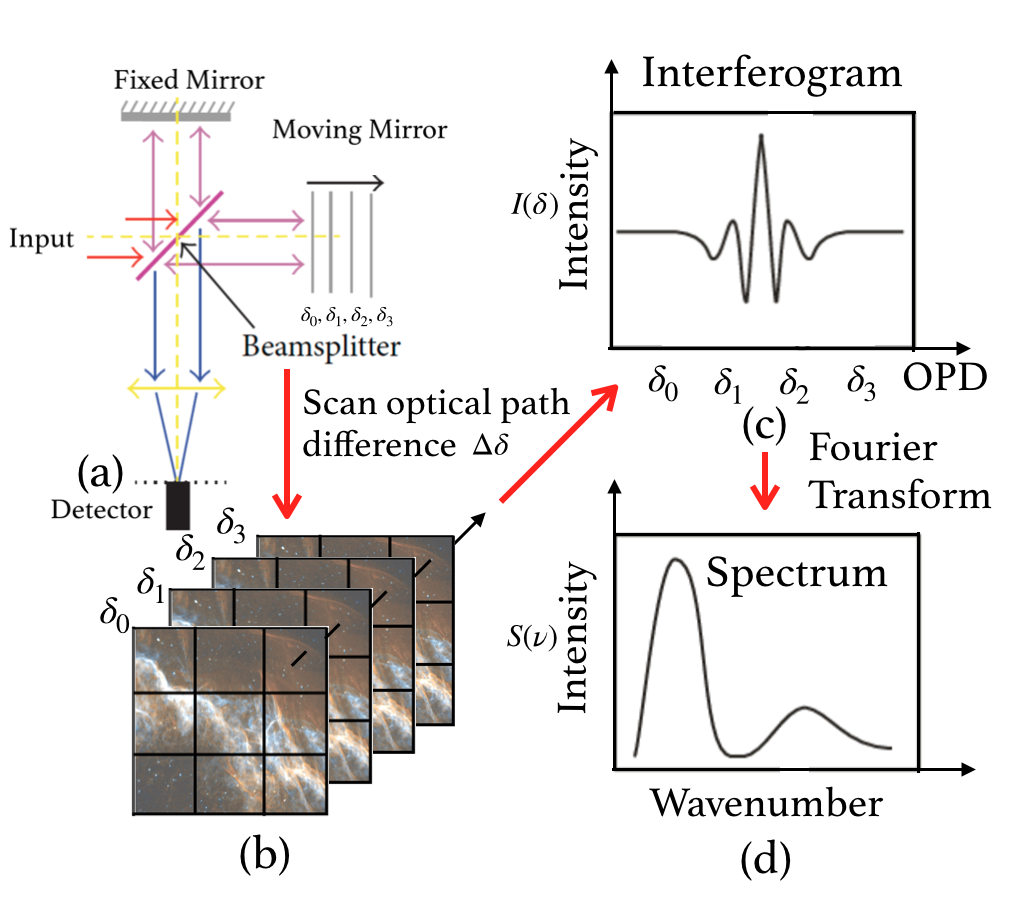}
    \caption{The data acquisition of an iFTS. The core optical configuration of an iFTS is a Michelson interferometer with a fixed mirror and a movable mirror. By moving the position of one mirror and taking images at every step, one can get the interferogram for every pixel in the image as a function of optical path difference. The measured interferogram contains all spectral information and can be Fourier transformed to obtain the spectrum of the input light. Panel (a) is adapted from \cite{Drissen2014}.}
    \label{fig:figure1}
\end{figure}

An imaging Fourier transform spectrograph (iFTS) computes the autocorrelation \textit{temporally}, by moving a mirror \citep{kauppinen2001fourier}.  In its classic one-port implementation, an iFTS is essentially a Michelson interferometer, consisting of a beamsplitter, fixed mirror, moving mirror, and imaging optics ( Figure~\ref{fig:figure1}). For an ideal beamsplitter, the incident beam is divided into two equal parts that go toward two mirrors and are reflected back. The path length of each beam is twice of the length of each arm, and a translation of the moving mirror by $\frac{\delta}{2}$ creates an optical path difference of $\delta$ between the two beams when they recombine and interfere. Therefore, by moving the position of one mirror, the optical path difference between the two beams changes, leading to a modulation of their interfered intensity at the focal plane. Retrieving spectral information involves scanning the optical path difference in discrete steps and recording the interfered intensity on the focal plane at every step, then taking a Fourier transform of the result.

To examine this in detail, consider two electromagnetic waves with electric fields  $E_{1}=A_{1}e^{i(kz-\omega t)}$ and $E_{2}=A_{2}e^{i(kz-\omega t+\phi)}$ propagating in the same direction $z$. Here $A_{i}$ is the amplitude of the wave, $k$ is the wave vector, $z$ is the position, $\omega$ is the angular frequency, $t$ is time, and $\phi$ is the phase difference between two waves. If $I_{1}$ and $I_{2}$ are the intensities of the two waves, the intensity of the interfered wave is $I=I_{1}+I_{2}+2\sqrt{I_{1}I_{2}} \cos \phi$. For the case of a classic one-port imaging Fourier transform spectrograph, if the input light intensity is $I_{0}$ and with an ideal beamsplitter ($50\%$ transmissivity and $50\%$ reflectivity), the output will be the inference of two waves with equal intensity $I_{1}=I_{2}=\frac{I_{0}}{4}$ as input light passes through the beamsplitter twice. Thus intensity of resulting wave can be expressed:
\begin{equation}
    I=\frac{I_{0}}{2}(1+\cos \phi)
\end{equation}
Assuming the spectrum of input light is continuous, it can be described as a function of wavenumber $B(\nu)$, where $\nu=\frac{1}{\lambda}$ is the wavenumber, with units of waves per unit distance. It is convenient to present the spectrum with respect to wavenumber rather than wavelength, with $B(\lambda)=\nu^{2}B(\nu)$. 

The interference signal from the whole spectral band is
\begin{equation}
    I = \int_{0}^{\infty} \frac{B(\nu)}{2} (1+cos\phi)d\nu
\end{equation}

The phase difference $\phi$ is determined by the optical path difference $\delta$ between two beams $\phi=\frac{2\pi \delta}{\lambda}=2\pi \nu \delta$, thus the interfered  intensity as a function of optical path difference is

\begin{equation}\label{equ:e3}
    I(\delta) = \int_{0}^{\infty} \frac{B(\nu)}{2} (1+cos(2\pi \nu \delta))d\nu
\end{equation}

\begin{figure}
    \centering
    \includegraphics[width=\linewidth]{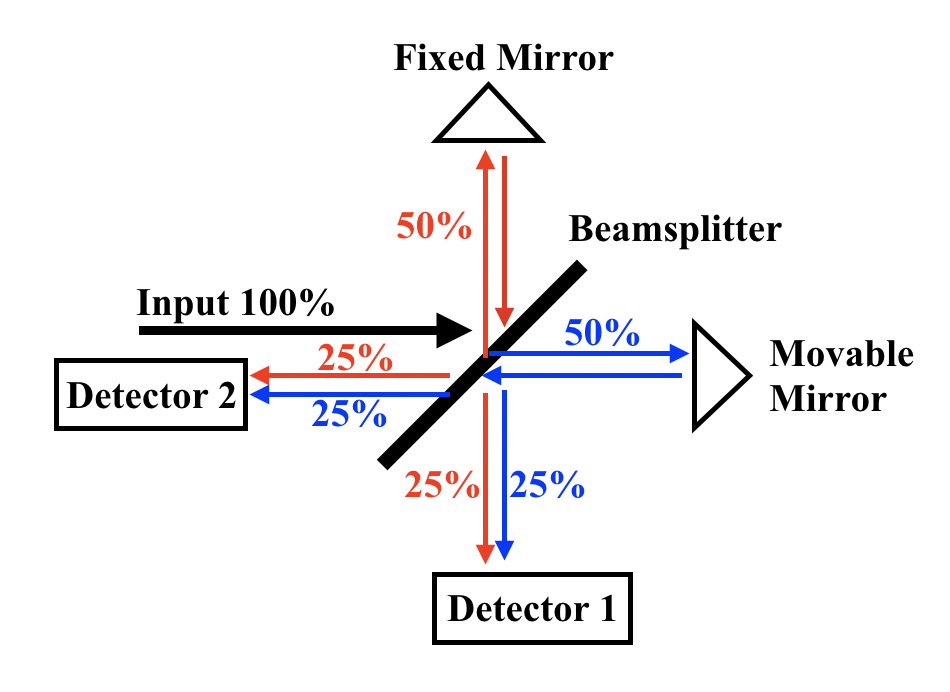}
    \caption{Schematic optical layout of a two-port iFTS with an ideal beamsplitter ($50\%$ transmissivity and $50\%$ reflectivity). The light passes through the beamsplitter twice: the input light first splits into two equal beams that go toward two mirrors. After being reflected by the mirrors, the two beams pass through the beamsplitter again and each of them is split again. The input light all goes to one of two detectors, ignoring the light loss from optical elements.}
    \label{fig:figure2}
\end{figure}

Equation~(\ref{equ:e3}) is for a classical one-port iFTS which detects half of the incident light, on average, as the other half goes back to the source after the second pass through the beamsplitter. Modern implementations use a more efficient two-port iFTS (see Figure~\ref{fig:figure2}), which directs the output beam to two cameras \citep{DavisBook2001}. In this design, every photon that enters the instrument is detected in one of these two output ports, excluding efficiency losses.  A dual-port iFTS theoretically doubles the amount of photons the instrument collects with respect to a single-port interferometer.  But the detector noise is also increased since the system requires two cameras. The recorded intensities at two outputs are anti-phased:

\begin{equation}
    I_{\pm}(\delta) = \int_{0}^{\infty} \frac{B(\nu)}{2} (1\pm \cos(2\pi \nu \delta))d\nu
\end{equation}
\noindent where the $\pm$ signs correspond to the two output ports. The sum and difference of the two interference signals at two ports are 

\begin{equation}
    I_{\sum}(\delta) =I_{+}(\delta)+I_{-}(\delta) =\int_{0}^{\infty} B(\nu) d\nu
\end{equation}

\begin{equation}\label{eq:po}
    I_{\Delta}(\delta) =I_{+}(\delta)-I_{-}(\delta)=\int_{0}^{\infty} B(\nu) cos(2\pi \nu \delta) d\nu
\end{equation}

The sum of the two port signals $I_{\sum}(\delta)$ is the integral of the full band intensity by conservation of energy, which is equivalent to the broadband (i.e. defined by any upstream filter) image of the targets. The difference signal $I_{\Delta}(\delta)$ is called the \textit{interferogram}, and can be used to obtain the spectrum. Therefore, one gets a concurrent image (by summing the two port signals) and spectrum (by subtracting one port signal from another) with a two-port iFTS. 

\begin{figure}
    \centering
    \includegraphics[width=\linewidth]{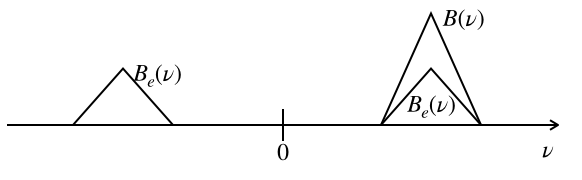}
    \caption{Construction of a symmetric spectrum $B_{e}(\nu)$ from an asymmetric spectrum $B(\nu)$ for Fourier transform. }
    \label{fig:figureBBe}
\end{figure}

 In practice, spectra only exist in the positive wavenumber range. However, the operation of the Fourier transform integrates the spectrum over all wavenumbers from $-\infty$ to $\infty$. Thus, following \citet{DavisBook2001}, it is useful to construct a symmetric spectrum $B_{e}(\nu)$ from the asymmetric spectrum $B(\nu)$ (see Figure~\ref{fig:figureBBe}):
\begin{equation}
    B_{e}(\nu)=\frac{1}{2}[B(-\nu)+B(\nu)].    
\end{equation}
The integral of $B(\nu)$ over positive wavenumbers  is equivalent to the integral of $B_{e}(\nu)$ over negative and positive wavenumbers. Because the cosine function is also even, Eqn.~\ref{eq:po} may be modified to
\begin{equation}
    I_{\Delta}(\delta) =\int_{-\infty}^{\infty} B_{e}(\nu) cos(2\pi \nu \delta) d\nu 
\end{equation}

$I_{\Delta}(\delta)$ and $B_{e}(\nu)$ form a Fourier transform pair. \footnote{Technically, Eq.7 $\&$ 8 present a cosine transform pair. However, due to the interferogram being a real-valued signal and the nearly negligible imaginary component of the spectrum when the interferogram exhibits symmetry, we refer it to as a Fourier transform pair.} Fourier transforming the interferogram $I_{\Delta}(\delta)$ yields the spectrum:

\begin{equation}\label{eqn:E_nu}
    B_{e}(\nu)= \int_{-\infty}^{\infty} I_{\Delta}(\delta) \cos(-2\pi \nu \delta) d\delta
\end{equation}

The Fourier transform of interferogram distributes the intensity into both positive and negative spectral domain. The output, symmetric spectrum $B_{e}(\nu)$ has an amplitude half that of the original input spectrum $B(\nu)$. 
 
 The above derivations hold for the interferograms of each pixel in the frame. If imaging optics are used to create focal planes at the two output ports, each pixel will measure an interferogram, with independent spectral information corresponding to the different points on the sky. In this way, an iFTS is able to retrieve spatially resolved spectral information from a two-dimensional focal plane.

Practical operation of an iFTS involves choices about the scan length, mirror step size, and time spent per step, which affect the resolving power and signal-to-noise ratio of the resulting spectrum. We direct the reader to the appendix~\ref{sec:POS} for a more detailed discussion of these issues.

\section{Methodology}\label{sec:me}
\subsection{Detector noise model} \label{sec:AF}

Detector noise is of primary importance in space-based exoplanet imaging and especially spectroscopy. In this work, we consider three types of detector noise: dark current, readout noise and clock-induced charge, ignoring some more complex effects such as charge trapping, cosmic ray amplification, and dynamic range compression. Dark current has a Poisson distribution with a variance per hour as $\sigma_{DC}^2 = DC$, where $DC$ is the dark current rate in number of electrons per pixel per hour.  Read noise is Gaussian with a variance given by the single-frame readout noise $RN^2$ (in electrons per pixel per frame) multiplied by the number of frames, $\sigma^2_{RN} = RN^2 N_{\rm{frm}}$ following the fact that $var(\sum _N X_i) = N var(X_i)$. The number of frames $N_{\rm{frm}}$ per hour can be approximated by $1/t_{\rm{frm}}$ where $t_{\rm{frm}}$ is the integration time per frame in hours. Therefore, the read noise variance per hour is $\sigma^2_{RN} = \frac{RN^2}{\rm{t_{frm}}}$. Similarly, clock-induced charge has a Poisson distribution with a variance per hour equal to the clock-induced charge $CIC$ (in electrons/pixel/frame) multiplied by the number of frames per hour, $\sigma^2_{CIC} = \frac{CIC}{t_{\rm{frm}}}$. Combining these terms, the detector noise variance per pixel per hour can be expressed as
\begin{equation}
    \sigma^{2}_{d} = DC + \frac{RN^{2}}{t_{\rm{frm}}} + \frac{CIC}{t_{\rm{frm}}}
\end{equation}

Unfortunately, it is not possible to eliminate the effects of read noise and clock-induced charge by taking very long frames, due to cosmic ray saturation. The above computation of the ``variance rate'' has been used in previous studies \citep{Robinson2016,Lacy2019,LUVOIRTeam} for IFS instruments, and we adopt it for the iFTS as well.  We use $t_{\rm{frm}}$ of 100  and 1000 seconds in our simulation for both the IFS and iFTS, which are approximately the maximum exposure times baselined for the \textit{Roman-CGI} and \textit{JWST}.

The breakdown of noise into dark and clock-induced charge is most relevant to EMCCD-based coronagraphs, such as the Roman CGI.  Recent work on single-photon sensitive and photon number resolving optical detectors like Skipper CCDs \citep{tiffenberg2017single} and scientific CMOS sensors \citep{gallagher2022characterization} can have negligible CIC, but may still possess dark current.  Future work is needed to assess these sensors' suitability for space-based high contrast imaging systems.

The number of pixels involved in the spectral observations is another factor determining the noise. As an IFS and iFTS employ different operational principles, they have different numbers of pixels contributing to an observation. For an IFS, spatially resolved spectroscopic observations are obtained by placing a grid of lenslets in the focal plane which transfer the light to dispersive elements (e.g. prisms or gratings). The light from each lenslet is spectrally dispersed and recorded on different pixels. The number of contributing pixels per spectral element is the product of the number of lenslets $n_{\rm{len}}$ involved and the number of pixels per spectral element per lenslet $n_{\rm{pix}}$. 

\begin{figure}
    \centering
    \includegraphics[width=\linewidth]{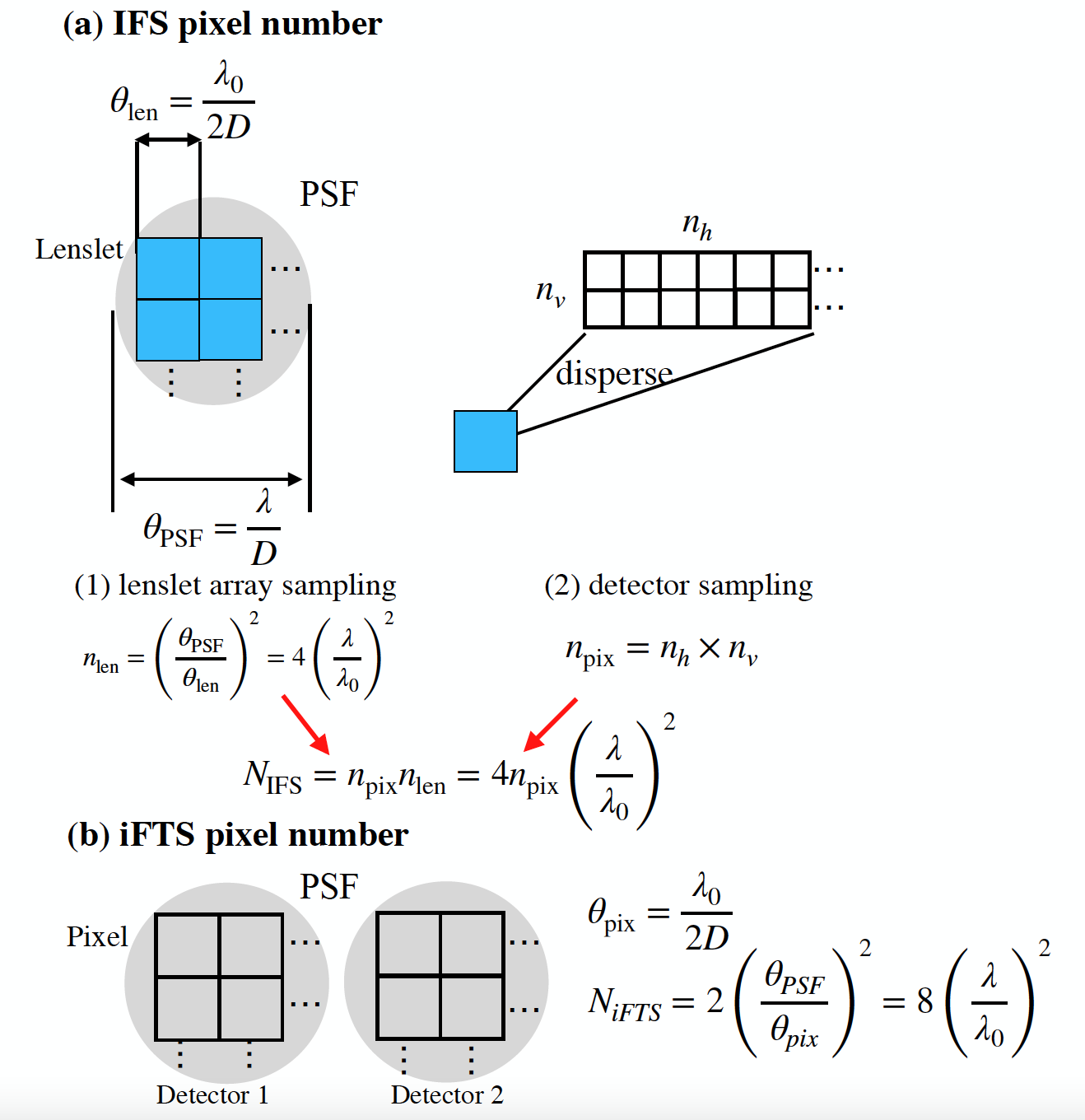}
    \caption{Schematic illustration showing the number of pixels contributing to one spectral element of an IFS and an iFTS. (a) An IFS works in the way of placing a lenslet array in the focal plane, which transfer the light from individual location to the prims or gratings. The prims or gratings disperse light into different pixels to get spectral information. The number of pixel contributing to an spectral element equals to the number of lenslets inside PSF area $N_{len}$ multiplied by the number of pixels per spectral element per lenslet $n_{pix}$. (c) An iFTS does not disperse photons so that the number of pixels contributing to each spectral elements is those inside the PSF area. Because an two port iFTS uses two cameras, we multiply the pixel number by two.    }
    \label{fig:figure15}
\end{figure}

Assuming the lenslet array is Nyquist sampled at the wavelength $\lambda_{0}$ (see Figure~\ref{fig:figure15}),the angular diameter of the lenslet $\theta_{\rm{len}}$ is
\begin{equation}\label{eqn:nyquistlenslet}
    \theta_{\rm{len}}=\frac{\lambda_{0}}{2D}
\end{equation}
where $D$ is the telescope diameter. We follow the optical design in LUVOIR report to Nyquist sample the lenslet array at the shortest wavelength of visible and NIR bands ($\lambda_{0}=540 nm$ in visible and $\lambda_{0}=1000 nm$ in NIR). At wavelength $\lambda$, the angular diameter of the PSF is $\rm{\theta_{PSF}=\lambda/D}$. The number of lenslets inside the point-spread function (PSF) at wavelength $\lambda$ is proportional to the square of the wavelength: 
\begin{equation}
    n_{\rm{len}} = \left(\frac{\theta_{\rm{PSF}}}{\theta_{\rm{len}}}\right)^{2}=4\left(\frac{\lambda}{\lambda_{0}}\right)^{2}
\end{equation}

Therefore, at the  Nyquist wavelength $\lambda_{0}$, there are four lenslets within the PSF, and this number increases at longer wavelengths.

The prisms or gratings disperse photons into different pixels. The number of pixels contributing to one spectral element for each lenslet $n_{pix}$ is the product of pixels in horizontal direction $n_{h}$ and vertical direction $n_{v}$; $n_{pix}=n_{h}\cdot n_{v}$. Thus the number of pixels contributing to one spectral element at wavelength $\lambda$ is 
\begin{equation}
    N_{\rm{IFS}} = n_{\rm{pix}}n_{\rm{len}}
    =4 n_{\rm{pix}}\left(\frac{\lambda}{\lambda_{0}}\right)^{2}
\end{equation}

The detector noise variance per spectral element per hour at wavelength $\lambda$ is then 
\begin{equation}\label{eq:sigmaifs1}
    \sigma^{2}_{\rm{IFS}} =\sigma^{2}_{d}\times N_{\rm{IFS}}
    =4\sigma^{2}_{d} n_{pix}\left(\frac{\lambda}{\lambda_{0}}\right)^{2}
\end{equation}

On the other hand, an iFTS does not disperse photons, so the number of contributing pixels is determined by the size of PSF aperture. If a pixel has an angular diameter $\theta_{\rm{pix}}$ and the detector is Nyquist sampled at wavelength $\lambda_{0}$, then 
\begin{equation}
    \theta_{\rm{pix}}=\frac{\lambda_{0}}{2D}
\end{equation}
where we use the same Nyquist wavelength $\lambda_{0}$ as the IFS (Eqn. \ref{eqn:nyquistlenslet}). Because a two-port iFTS uses two cameras, it requires twice the contributing pixels:
\begin{equation}
    N_{\rm{iFTS}}= 2\left(\frac{{\rm{\theta_{PSF}}}}{\rm{\theta_{pix}}}\right)^{2} =8\left(\frac{\lambda}{\lambda_{0}}\right)^{2}
\end{equation}

The detector noise in this case is introduced at the interferogram. The detector noise variance per hour at wavelength $\lambda$ is
\begin{equation}\label{eq:sigmaifts1}
    \sigma^{2}_{\rm{iFTS}}=\sigma^{2}_{d}\times N_{\rm{FTS}} =8\sigma^{2}_{d} \left(\frac{\lambda}{\lambda_{0}}\right)^{2}
\end{equation}

From the ratio of Eqn.'s \ref{eq:sigmaifs1} and \ref{eq:sigmaifts1}, it follows that  $\sigma^{2}_{\rm{IFS}} /\sigma^{2}_{\rm{iFTS}} = n_{pix}/2$, but this is not the end of the story, as the noise in the iFTS spectrum needs to be determined, as opposed to the noise in the interferogram. We will present the details about the noise propagation in frequency domain for iFTS in Section~\ref{sec:ftsmodel}, and the comparison of IFS and iFTS noise variance per spectral element in Section~\ref{sec:CMW}.  

\subsection{iFTS Simulation}\label{sec:ftsmodel}

In section~\ref{sec:FTSconcept}, we consider the spectrum and interferogram as continuous function and the Fourier transform is expressed as integrals from $-\infty$ to $\infty$. In practice, the measurement of a signal usually gives us a finite number of data measured at discrete points. Consequently, the integral from $-\infty$ to $\infty$ are replaced by a sum of discrete points from $-N$ to $N$, which is called the \textit{discrete Fourier transform}. Here, we will present our numerical simulations of these digital Fourier transforms and discuss the signal-to noise ratio model. 

Let $S(\nu_{i})$ denote the discrete input spectrum coming into the iFTS, where the wavenumber $\nu_{i}=i\Delta \nu$, for $i =0,1,2...N$. As discussed in section~\ref{sec:FTSconcept}, it is convenient to use the symmetric spectrum $S_{e}(\nu_{i})=\frac{1}{2}[S(\nu_{i})+S(-\nu_{i})]$, where $i =-N,-N+1...-1,0,1,...N$. The units for  $S(\nu_{i})$ and  $S_{e}(\nu_{i})$ are photon counts per wavenumber per hour. 

Based on Eqn. 8, the interferogram is obtained from the difference between intensities at two cameras:
\begin{figure*}
    \centering
    \includegraphics[width=0.9\linewidth]{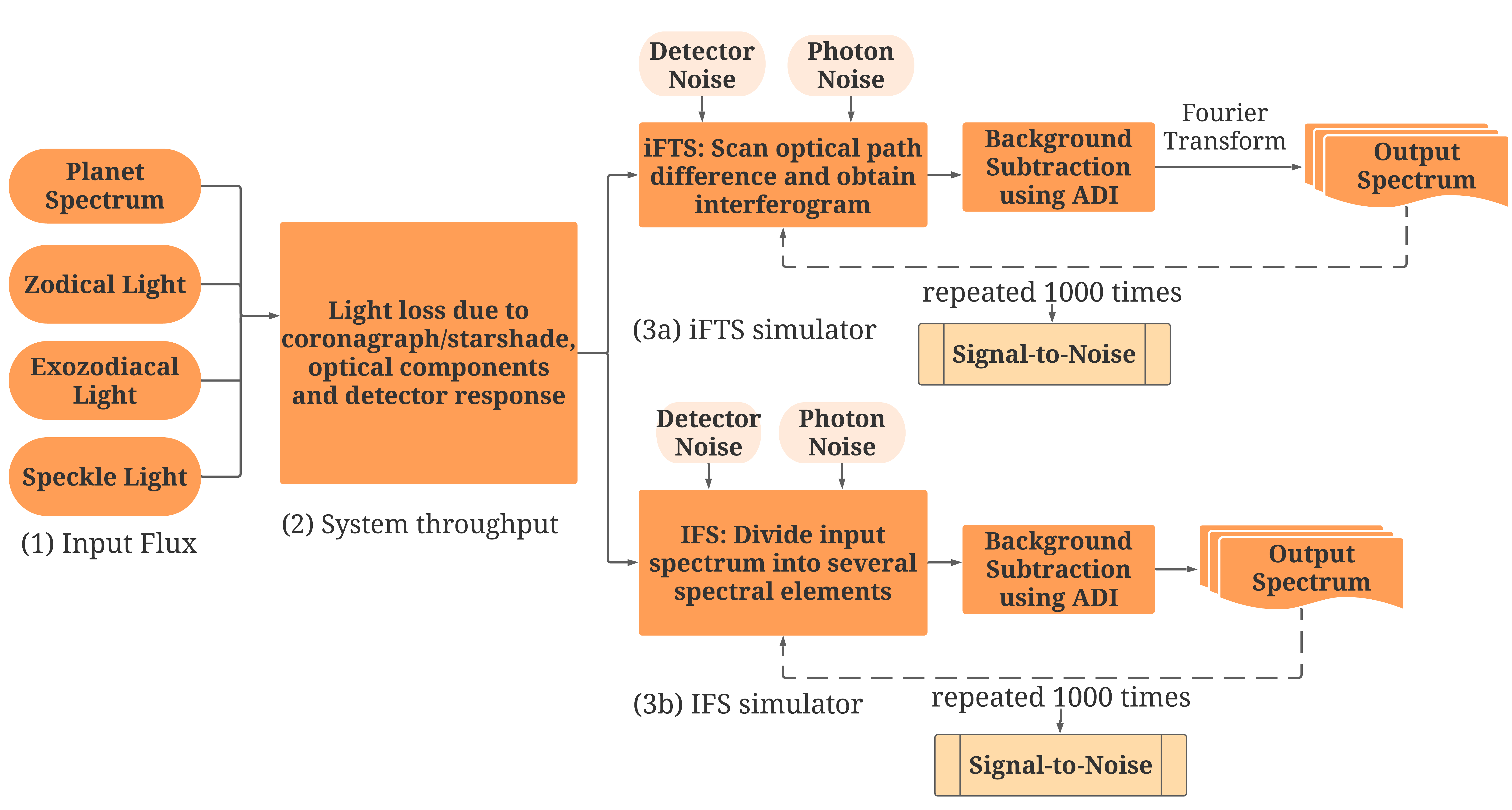}
    \caption{Flow chart of the numerical simulation.}
    \label{fig:figureaadd1}
\end{figure*}
\begin{equation}\label{eqn:inf}
\begin{split}
   F_{\Delta}(\delta_{j}) & =F_{+}(\delta_{j})-F_{-}(\delta_{j}) \\
   & = T(\delta_{j})\sum_{i=-N}^{N}{S_{e}(\nu_{i})\cos{(2\pi \nu_{i} \delta_{j})}}\Delta \nu
   \\
   \delta_{j}& =j\Delta \delta, j=-N,-N+1..,-1,0,1...N-1,N
\end{split}
\end{equation}

where $\delta_{j}$ is the optical path difference between the two beams (with a unit of nm) and $T(\delta_{j})$ is the exposure time at every scanning step (with a unit of hour). The simulated interferogram $F_{\Delta}(\delta_{j})$ has a unit of photon counts.  The interferogram is double-sided and symmetric about the zero optical path difference. We also add photon noise and detector noise to the intensity at two cameras at every step. The difference of intensity at two cameras has the same noise variance as the sum of intensity, which is the the total intensity of the incident light during the exposure time (photon noise) and the detector noise.
 
\begin{equation}\label{eqn:k}
\begin{split}
    var[F_{\Delta}(\delta_{j})] &= var[F_{\sum}(\delta_{j})]\\
    &=T(\delta_{j})\sum_{i=-N}^{N} {S_{e}(\nu_{i})\Delta \nu +  \sigma^{2}_{\rm{iFTS}}}T(\delta_{j})
\end{split}
\end{equation}
where $var[X]$ represents the variance of the variable X. 

We then apply the Fourier transform to the simulated interferogram $ F_{\Delta}(\delta_{j})$. Based on Eqn.~\ref{eqn:E_nu}, the output spectrum will be a symmetric spectrum in both negative and positive frequencies.  \footnote{The imaginary part is nearly zero because the interferogram is double sided and symmetric}.
\begin{equation}\label{eqn:ftsout}
    O_{\rm{FTS}}(\nu_{i}) = \sum_{j=-N}^{N} F_{\Delta}(\delta_{j}) cos(-2\pi \nu_{i} \delta_{j})\Delta \delta 
\end{equation}

 The scans only cover a finite range of the interferogram instead of from $-\infty$ to $\infty$. The truncation of the interferogram in a range of $(-L,L)$ leads to the convolution of the original spectrum with an Instrument Line Shape (ILS) function. If we assign a total exposure time $T_{tot}=\sum_{-N}^{N} T(\delta_{j})$ evenly at every step, the truncated interferogram can be seen as the product of a infinite interferogram with a boxcar function (see details in Appendix~\ref{sec:resolution}). Thus, the output spectrum is the convolution of the $S_{e}(\nu_{i})$ with a sinc function times the total exposure time
\begin{equation}\label{eq:signal}
    O_{\rm{FTS}}(\nu_{i}) = T_{tot} [S_{e}(\nu_{i}) \circledast 2L sinc(2L\nu_
{i})] 
\end{equation}

If the full width at half maximum (FWHM)  of the ILS is $\Delta \nu_{\rm{ILS}}$, the intensity of output spectrum will be 
\begin{equation}\label{eq:signal}
    O_{\rm{FTS}}(\nu_{i}) =S_{e}(\nu_{i}) \Delta \nu_{\rm{ILS}} T_{\rm{tot}}
\end{equation}

where the output spectrum has a unit of photon counts per spectral element (in total time).  Based on the Rayleigh energy theorem, the total noise variance in spectral domain equals to the total noise variance in the interferogram domain. But as our computation only involves the cosine terms of the Fourier transform, the variance of the output spectrum is only half of the total variance of the interferogram (for details see chapter 8 in \citealt{DavisBook2001}).

\begin{equation}\label{eq:ss}
   \frac{1}{2N}\sum_{-N}^{N} var[O_{\rm{FTS}}(\nu_{i})]
   = \frac{1}{2}\sum_{-N}^{N} var[ F_{\Delta}(\delta_{j})]   
\end{equation}

As the spectral noise is white and independent of the wavenumbers,  the sum sign and $2N$ in the denominator in the left term of Eqn. ~\ref{eq:ss} can be canceled out and we have

\begin{equation}\label{eq:ss2}
\begin{split}
   var[O_{\rm{FTS}}(\nu_{i})] 
   & = \frac{1}{2}\sum_{-N}^{N} var[ F_{\Delta}(\delta_{j})]\\
   & =\frac{T_{\rm{tot}}\sum_{-N}^{N} {S_{e}(\nu)\Delta \nu +  \sigma^{2}_{\rm{iFTS}}}T_{\rm{tot}}}{2}
\end{split}
\end{equation}
In the numerator of Eqn.~\ref{eq:ss2}, the first term represents the total photon count entering the iFTS during the entire scan (photon noise), while the second term accounts for the total detector noise variance. Combining Eq~\ref{eq:signal} and Eq~\ref{eq:ss2}, the signal-to-noise ratio of the output spectra can be described by 
\begin{equation}
    \rm{SNR_{FTS}(\nu_{i})}= \frac{S_{e}(\nu_{i}) \Delta \nu_{\rm{ILS}} T_{\rm{tot}}}{\sqrt{\frac{1}{2}[T_{\rm{tot}}\sum_{-N}^{N} {S_{e}(\nu_{i})\Delta \nu +  \sigma^{2}_{\rm{iFTS}}}T_{\rm{tot}}]}}
\end{equation}
The above expression is a symmetric function from negative to positive wavenumbers. If we go back to the convention of the input spectrum $S(\nu_{i})$ only spanning in positive frequencies, we then obtain

\begin{equation}
    \rm{SNR_{FTS}(\nu_{i})}= \frac{S(\nu_{i}) \Delta \nu_{\rm{ILS}} T_{\rm{tot}}}{\sqrt{2[T_{\rm{tot}}\sum_{0}^{N} {S(\nu_{i})\Delta \nu +  \sigma^{2}_{\rm{iFTS}}}T_{\rm{tot}}]}}
\end{equation}


In above equations, $S(\nu_{i})$ denotes the input signal fed into the iFTS, which in these simulations consists of the planet spectrum $c_{pl}(\nu_{i})$ and Solar System zodiacal light $c_{z}(\nu_{i})$, exozodiacal light $c_{ez}(\nu_{i})$ and residual starlight $c_{sp}(\nu_{i})$ in unit time (details see Appendix~\ref{sec:plab}). We assume stellar flux is suppressed by a factor $C$ at the image plane after a coronagraph, e.g. $C$ is $10^{-10}$ for the Habitable World Observatory. We also consider light loss due to the finite throughput of coronagraphs $\tau_{core}$, spectrograph optics $\tau_{optical}$ and the quantum efficiency of the detectors $\tau_{\rm{QE}}$ (details see Appendix~\ref{sec:ST}). We use the same optical throughput for an iFTS and an IFS for simplicity, though the FTS may be expected to have slightly higher throughput due to its purely reflective design. Figure~\ref{fig:figureaadd1} shows a flow chart to illustrate the procedure and workflow of our numerical simulations. 

Practical observations will involve background subtraction of some sort, to remove systematic errors from speckles, zodi, exozodi, and detector noise. In this case, we assume the observation uses the technique of roll deconvolution or angular differential imaging (ADI,\citealt{Marois2006}), where the target star is observed at two roll angles, allowing for background subtraction with the planet at two different positions (we assume smooth zodi and exozodi backgrounds).  The effect of this subtraction is to increase the background variance by a factor of two:\footnote{ Other background subtraction techniques using reference stars or models can result in smaller noise increases, but these involve assumptions about the targets and the ability to model systematic errors \citep{Lacy2019}.}

\begin{equation}\label{eqn:snr}
   \rm{SNR_{FTS}^{'}(\nu_{i})}= \frac{c_{pl}(\nu_{i}) \Delta \nu_{\rm{ILS}} T_{\rm{tot}}}{\sqrt{2T_{\rm{tot}}[ \sum_{0}^{N} c_{pl}(\nu_{i}) \Delta \nu +2 \sum_{0}^{N} c_{b}(\nu_{i})\Delta \nu + 2 \sigma^{2}_{\rm{iFTS}}}]}
\end{equation}
where we use $c_{b}(\nu_{i}) = c_{z}(\nu_{i})+c_{ez}(\nu_{i})+c_{sp}(\nu_{i})$ to denote the total astrophysical background rate. In this way, we can get the detector noise variance per spectral element in exposure time $\mathrm{T_{tot}}$ for iFTS is $var[D_{\mathrm{iFTS}}]=4\sigma^{2}_{\mathrm{iFTS}}\mathrm{T_{tot}}= 32\sigma^{2}_{d}(\frac{\lambda}{\lambda_{0}})^{2}\mathrm{T_{tot}}$.

Theoretical derivations of SNRs in Fourier transform spectrographs are provided in \citet{Treffers1977, DavisBook2001, Graham1998} and \citet{bennett1999critical}. 
They achieved consistent analytical models when using a double-sided interferogram. But the latter two claim  a better SNR by a factor of $\sqrt{2}$ with a single-sided interferogram, which are inconsistent with the former two. To address this discrepancy, we carried out Monte Carlo simulations of the FTS data acquisition and transformation process, resulting in the posterior distribution of intensity for each spectral element. The numerical SNR is computed as the mean of the posterior distribution divided by its standard deviation. Fig.~\ref{fig:figure5} exhibits two such sample simulations.  The simulated SNRs are found to be in very good agreement with the theoretical predictions of \citet{Treffers1977} and \citet{DavisBook2001}, but do not yield the extra $\sqrt{2}$ noise reduction predicted by \citet{Graham1998} and \citet{bennett1999critical}.

\begin{deluxetable}{lp{3cm}c}[b]

\tablecaption{Summary of Astrophysical Parameters\label{tab:rvs}
}
\tablehead{\colhead{Parameter} & \colhead{Description} & \colhead{Fiducial Value} }
\startdata
 & Name  & Earth twin \\
$R_{p}$ & planet radius  & $1\ R_{\oplus}$ \\
$a$ & planet orbit distance  & 1 AU   \\
$d$ & system distance from Earth  & 10 pc   \\
$\alpha$ & planet orbit phase angle  & $60^{\circ}$   \\
$T_{\star}$ & temperture of the host star  & 5770 K  \\
$F_{o,V}$ & standard zero-magnitude V-band specific flux & $3.63\times 10^{-11} \rm{W\ m^{-2}\  nm^{-1}}$  \\
$F_{\odot,V}(1AU)$ & solar flux density at 1 AU in V band &  $1.86\times 10^{3} \rm{W\ m^{-2}\ nm^{-1}}$ \\
$M_{z,V}$ & V-band zodiacal light surface brightness & $23\ \rm{mag\ arcsec^{-2}}$  \\
$M_{ez,V}$ & V-band exozodiacal light surface brightness & $22\ \rm{mag\ arcsec^{-2}}$  \\
$N_{ez,V}$ & number of exozodis in exoplanetary disk & 1 \\
\enddata


\end{deluxetable}

\subsection{IFS Simulation}\label{sec:ifsmodel}

For simulating the IFS performance, it is more convenient to work in units of wavelength rather than wavenumber, hence $S(\lambda)$ denotes the input spectrum.  We divide the input spectrum into several spectral elements and the width of each element depends on the resolving power $R$: $\Delta \lambda_{\rm{eff}} = \frac{\lambda}{R}$. We then integrate the spectrum in each element and add the detector noise and multiply them with the total exposure time $T_{\rm{tot}}$. The signal is:
\begin{equation}
\begin{split}
    O_{\rm{IFS}}(\lambda) & =  T_{tot} \int_{\lambda}^{\lambda+\Delta \lambda_{\rm{eff}}}S(\lambda) d\lambda \\
    & \approx S(\lambda) \Delta \lambda_{\rm{eff}} T_{tot}
\end{split}
\end{equation}

The noise variance of each element is the sum of  photon counts and detector noise
\begin{equation}\label{eq:IFSsigma}
    \sigma^{2}[O_{\rm{IFS}}] =  S(\lambda) \Delta \lambda_{\rm{eff}} T_{tot} + \sigma^{2}_{\rm{IFS}}T_{\rm{tot}}
\end{equation}

Therefore, the SNR for an IFS is 
\begin{equation}
    \rm{SNR}_{\rm{IFS}}=\frac{  S(\lambda) \Delta \lambda_{\rm{eff}} T_{tot} }{\sqrt{S(\lambda) \Delta \lambda_{\rm{eff}}  T_{tot} + \sigma^{2}_{\rm{IFS}}T_{\rm{tot}}}}
\end{equation}

When using ADI for background subtraction, the SNR is:
\begin{equation}\label{eqn:ifssnrfinall}
    \rm{SNR}_{\rm{IFS}}^{\prime}=\frac{c_{pl}(\lambda) \Delta \lambda_{\rm{eff}}  T_{tot} }{\sqrt{c_{pl}(\lambda) \Delta \lambda_{\rm{eff}} T_{tot} +2c_{b}(\lambda) \Delta \lambda_{eff}  T_{tot} +2\sigma^{2}_{\rm{IFS}}T_{\rm{tot}}}}
\end{equation}

\begin{figure*}
    \centering
    \includegraphics[width=\linewidth]{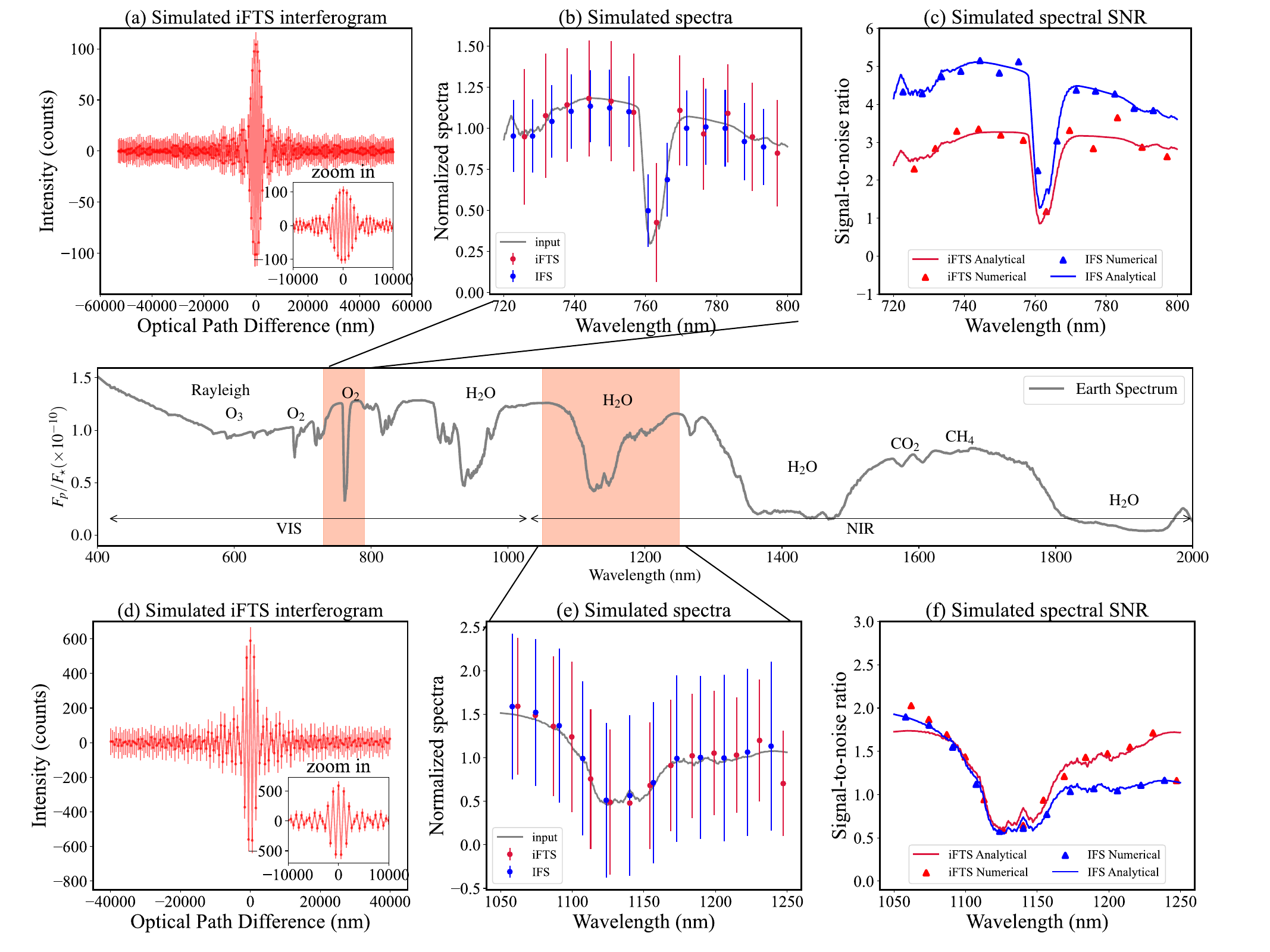}
    \caption{Examples of simulation results for an Earth twin around \textbf{a 4.83 mag G0V star} at 10 pc in a visible (710-800 nm, R=140, top) and NIR bandpass (1050-1250 nm, R = 70, bottom). (a)$\&$(d): simulated interferograms as the function of the optical path differences (OPDs) of a two-port iFTS. Red points are the difference of intensity at two outputs at each scanning step, with error bars from photon noise and detector noise. \textbf{We adopted a double-sided Boxcar apodization function, with constant exposure times assigned to every step and symmetric across zero-OPDs. The maxium OPDs are determined by the desired resolving power. } \textbf{The insert panels present a zoom-in around zero-OPDs.}  (b)$\&$(e) present simulated spectra for an iFTS (red) and an IFS (blue) \textbf{after background subtraction. The gray lines are normalized input planet spectra. The dots and error bars correspond to the mean value and standard deviation of 1000 simulations.} (c)$\&$(f) present the analytical and numerical signal-to-noise ratios of an iFTS (red) and an IFS (blue). \textbf{The numerical simulations produce consistent SNR with those from analytical models.}
     Here, we use a 6 m sized telescope with dark current of $0.77\ \rm{e^{-}/hr/pix}$ for the visible camera and readout noise of $5.9\ \rm{e^{-}/frame/pix}$ for the NIR camera with a frame time of 1000s. The SNRs are computed for a 500 hrs exposure time analytically and numerically. Other instrument parameters see in Table~\ref{tab:so} }
    \label{fig:figure5}
\end{figure*}
\begin{figure*}
    \centering
    \includegraphics[width=\linewidth]{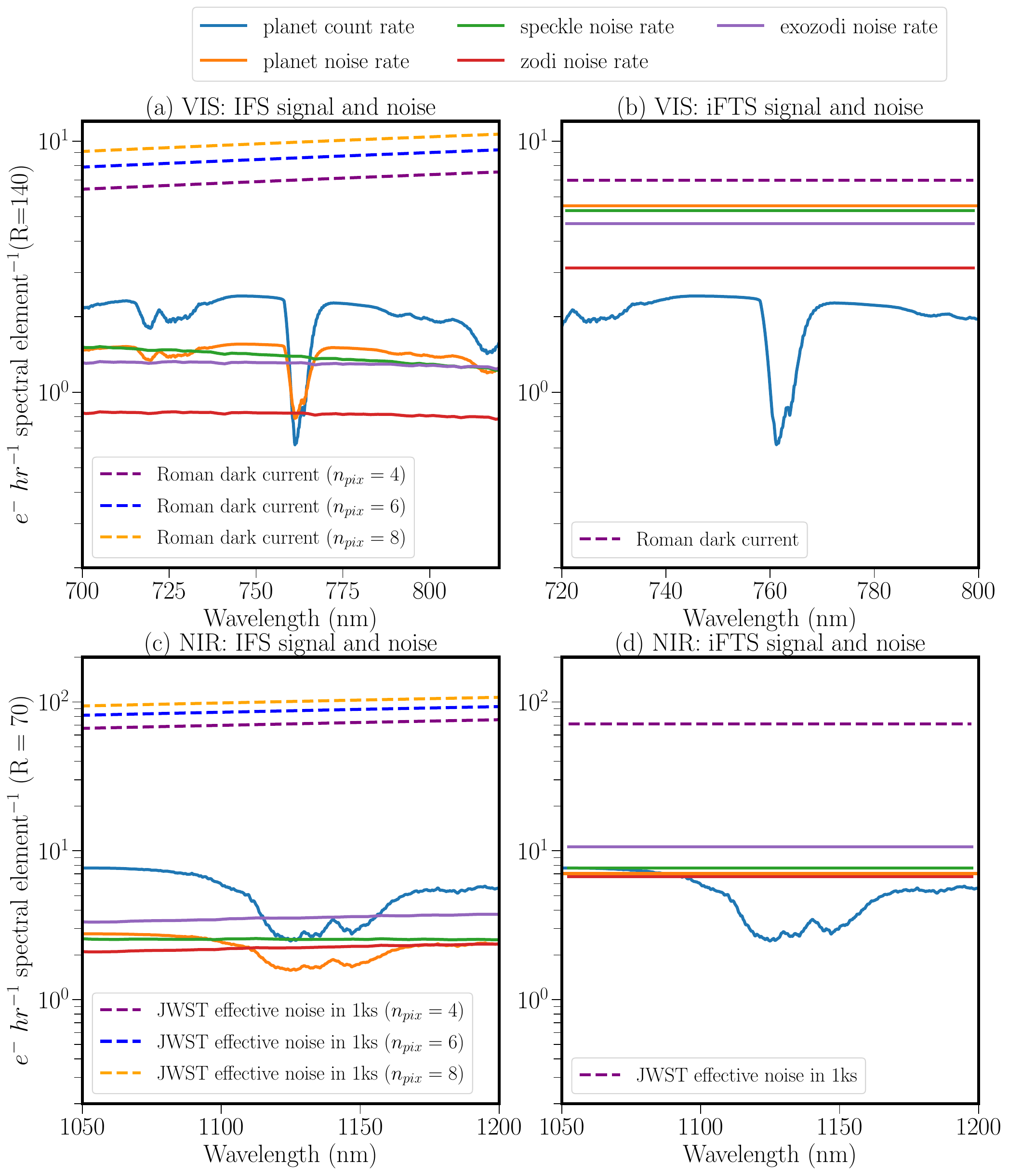}
    \caption{Comparison of plant count rate and major noise rates for the fiducial planet-star system using the following configurations: (a) IFS within the visible band (700-820 nm); (b) iFTS within the visible band (700-820 nm); (c) IFS within the NIR band (1050-1200 nm); (d) iFTS within the NIR band (1050-1200 nm). We assume the fiducial target to be an Earth twin \textbf{around a 4.83 mag G0V star at 1 AU} located at a distance of 10 pc ( details see Table~\ref{tab:rvs}). The solid lines depict the rate of planetary signal and photon noise from the planet, residual starlight, Solar System Zodiacal light, and exo-zodiacal light, as described in  Eqn.~\ref{eqn:snr} and Eqn.~\ref{eqn:ifssnrfinall}.\textbf{ The noise rates for an iFTS are determined by the total noise added to the interferogram summer over every scanning step. So they are white noise and independent on the wavelength. } The dashed lines represent the dark currents of the Roman EMCCD ($0.77\ \rm{e^{-}/hr/pix}$) in the visible band and the readout noise of HxRG on JWST ($5.9\ \rm{e^{-}/frame/pix}$) in the NIR band. In the case of IFS, we present the detector noise rate for various numbers of pixels assigned to each spectral element. Other instrument parameters see in Table~\ref{tab:so}}
    \label{fig:figure_n1}
\end{figure*}
The detector noise variance per spectral element in exposure time $\mathrm{T_{tot}}$ for an IFS is $var[D_{\mathrm{IFS}}]=2\sigma^{2}_{\rm{IFS}}\mathrm{T_{tot}}=8\sigma^{2}_{d}n_{\mathrm{pix}}(\frac{\lambda}{\lambda_{0}})^{2}\mathrm{T_{tot}}$. We compared our IFS simulations to those in \cite{Robinson2016} and \cite{Lacy2019} and obtained values for the planet photon flux and noise count rate matching theirs within $5\%$. Figure~\ref{fig:figure5} also shows that our IFS numerical simulations agree with the analytical models.

\subsection{Signal and noise comparison between iFTS and IFS}\label{sec:CMW}

Figure~\ref{fig:figure_n1} illustrates the signal rate from an Earth twin around a 4.83 mag G0V star at 1 AU from 10 pc away and primary noise count rates using a 6-meter telescope with an IFS and an iFTS, respectively. In the visible, the dominant contributor to noise for the IFS is the dark current, which surpasses the planet signal by nearly twofold when adopting the EMCCD dark current value from the \textit{Roman-CGI}. In contrast, the iFTS has higher photon noise in the visible band due to the integrated photon noise across the entire bandpass. The photon noise of iFTS is comparable to the detector noise in the visible channel, which makes it less effective than an IFS.

In the NIR, the readout noise is the most prominent noise source for both an IFS and an iFTS, exceeding the planet signal and photon noise by nearly one order of magnitude when using the  effective noise in 1000s of the HxRG detector in JWST. Therefore, the performance of an IFS and an iFTS are largely determined by the detector noise. In addition, the IFS detector noise also depends on the detector sampling design.  The ratio of detector noise variance per spectral element between an iFTS and an IFS is determined by the number of pixels allocated to an individual spectral element of the IFS, denoted as $n_{\rm{pix}}$:
\begin{equation}
    \frac{var[D_{\mathrm{IFS}}]}{var[D_{\mathrm{iFTS}}]}=\frac{n_{pix}}{4}
\end{equation}

The above expression implies that if an IFS uses 4 pixels per spectral element per lenslet, its detector noise will be almost the same as that of an iFTS. However, if using more than 4 pixels per spectral element per lenslet, an IFS will have higher detector noise than an iFTS. Note that setting $\mathrm{n_{pix}}=4$ corresponds to Nyquist sampling in both the vertical and horizontal directions on the CCD, using a $2\times 2$ grid.

If the two instruments have the same resolving power, then the required exposure time ratio between an IFS and an iFTS to achieve the same $\rm{SNR}$ is given by

\begin{equation}\label{eq:timer}
\frac{T_{\rm{IFS}}}{T_{\rm{iFTS}}} = \frac{c_{pl}(\lambda) \Delta \lambda_{\rm{eff}}  +2c_{b}(\lambda) \Delta \lambda_{\rm{eff}}  +2\sigma^{2}_{\rm{IFS}}} {  2[ \sum c_{pl}(\nu) \Delta \nu +2 \sum c_{b}(\nu)\Delta \nu + 2 \sigma^{2}_{\rm{iFTS}}] } 
\end{equation}

The required exposure time ratio is determined by total noise variance of the two spectrographs. 



\begin{deluxetable*}{llllllll}
\tablecaption{Simulation Overview}\label{tab:so}
\tablewidth{500pt}
\tabletypesize{\scriptsize}
\tablehead{\colhead{Num.$\&$Fig.} &\colhead{Simulation} &\colhead{$\tau_{\rm{core}}^{2}$}&\colhead{$\tau_{\rm{optical}}^{3}$} &\colhead{$n_{\rm{pix}}^{4}$} & \colhead{Resolving power} &\colhead{VIS detector noise$^{4}$}& \colhead{NIR detector noise$^{4}$} }
\startdata
\rule{0pt}{30pt} 
\makecell{Simulation1\\ Fig.6$\&$7} &\makecell{Example}   & $10\%$& \makecell{ $\rm{VIS}:24\%$\\ $\rm{NIR}:\ 32\%$}  &\makecell{ $\rm{VIS}:6$ \\ $\rm{NIR}:\ 4$}&\makecell{ $\rm{VIS}:140$ \\ $\rm{NIR}:\ 70$} & \makecell[l]{$\rm{i_{d}:0.77 e^{-}pix^{-1}hr^{-1}}$\\$i_{r}:0$\\ $\rm{q_{cic}:0.01 e^{-}pix^{-1}frame^{-1}}$\\$(t_{\rm{frm}}=100s)$ } &  \makecell[l]{$\rm{i_{d}:0}$\\$\rm{i_{r}: 5.9  e^{-}pix^{-1}frame^{-1}}$\\ ($t_{\rm{frm}}=1000s)$\\ $\rm{q_{cic}:0}$} \\
\rule{0pt}{30pt}
\makecell{Simulation2a\\ Fig.8} &\makecell{General \\in VIS}   & $10\%$& \makecell{ $\rm{VIS}:24\%$} &\makecell{ $\rm{VIS}:6$} &\makecell{ $\rm{VIS}:140$}&\makecell[l]{$\rm{i_{d}:0.1\sim3.4\  e^{-}pix^{-1}hr^{-1}}$\\$i_{r}:0$\\ $\rm{q_{cic}:0.01 e^{-}pix^{-1}frame^{-1}}$\\($t_{\rm{frm}}=100s)$ } &  --\\
\rule{0pt}{30pt}
\makecell{Simulation2b\\ Fig.9} &\makecell{General \\in VIS}   & $3\% - 40\%$& \makecell{ $\rm{VIS}:24\%$} &\makecell{ $\rm{VIS}:2-10$} &\makecell{ $\rm{VIS}:140$}&\makecell[l]{$\rm{i_{d}:0.1\sim3.4\  e^{-}pix^{-1}hr^{-1}}$\\$i_{r}:0$\\ $\rm{q_{cic}:0.01 e^{-}pix^{-1}frame^{-1}}$\\$(t_{\rm{frm}}=100s)$ } &  --\\
\rule{0pt}{30pt}
\makecell{Simulation3a\\ Fig.10} &\makecell{General\\ in NIR}  & $10\%$& \makecell{ $ \rm{NIR}:\ 32\%$} &\makecell{ $ \rm{NIR}:\ 6$} &\makecell{ $\rm{NIR}:\ 70$}&--&  \makecell[l]{$i_{d}:0 $\\$\rm{i_{r}: 0.1\sim 10 e^{-}pix^{-1}frame^{-1}}$\\ $(t_{\rm{frm}}=1000s)$\\ $\rm{q_{cic}:0}$} \\
\rule{0pt}{30pt}
\makecell{Simulation3b\\ Fig.11} &\makecell{General\\ in NIR}   & $3\%-40\%$& \makecell{ $ \rm{NIR}:\ 32\%$} &\makecell{ $ \rm{NIR}:\ 2\sim10$} &\makecell{ $\rm{NIR}:\ 70$}&--&  \makecell[l]{$\rm{i_{d}:0}$ \\ $\rm{i_{r}: 0.1\sim 10 e^{-}pix^{-1}frame^{-1}}$ \\ $(t_{\rm{frm}}=1000s)$\\ $\rm{q_{cic}:0}$} \\
\rule{0pt}{30pt}
\makecell{Simulation4a\\ Fig.12} &\makecell{General\\ $+$ varied \\noise }   & $10\%$ & \makecell{ $\rm{VIS}:24\%$\\ $\rm{NIR}:\ 32\%$} &\makecell{ $\rm{VIS}:6$ \\ $\rm{NIR}:\ 6$} & \makecell{ $\rm{VIS}:140$ \\ $\rm{NIR}:\ 70$}& \makecell[l]{$\rm{i_{d}:0.1\sim 3.4\ e^{-}pix^{-1}hr^{-1}}$ \\ $i_{r}:0$ \\ $\rm{q_{cic}: 0.01 \ e^{-}pix^{-1}frame^{-1}}$ \\ $(t_{\rm{frm}}=100s)$ }   &  \makecell[l]{$\rm{i_{d}:0}$\\$\rm{i_{r}:0.1\sim 10\ e^{-}pix^{-1}frame^{-1}}$\\$(t_{\rm{frm}}=1000s)$\\ $\rm{q_{cic}:0}$ } \\
\rule{0pt}{30pt}
\makecell{Simulation4a\\ Fig.13} &\makecell{HWO\\+ current detector \\noise }   &$10\%$& \makecell{ $\rm{VIS}:24\%$ \\ $\rm{NIR}:\ 32\%$} &\makecell{ $\rm{VIS}:6$ \\ $\rm{NIR}:\ 6$} &\makecell{ $\rm{VIS}:140$\\ $\rm{NIR}:\ 70$}& \makecell[l]{$\rm{i_{d}:0.77\  e^{-}pix^{-1}hr^{-1}}$\\$i_{r}:0$\\$\rm{q_{cic}: 0.01 \ e^{-}pix^{-1}frame^{-1}}$\\$(t_{\rm{frm}}=100s)$ }   &  \makecell[l]{$\rm{i_{d}:0}$\\$\rm{i_{r}:5.9\ e^{-}pix^{-1}frame^{-1}}$\\$(t_{\rm{frm}}=1000s)$\\ $\rm{q_{cic}:0}$ } \\
\rule{0pt}{30pt}
\makecell{Simulation4b\\ Fig.13} &\makecell{HWO \\+ future detector \\noise } & $10\%$ & \makecell{ $\rm{VIS}:24\%$ \\ $\rm{NIR}:\ 32\%$} &\makecell{ $\rm{VIS}:6$ \\ $\rm{NIR}:\ 6$} & \makecell{ $\rm{VIS}:140$ \\ $\rm{NIR}:\ 70$} &\makecell[l]{$\rm{i_{d}:0.1\  e^{-}pix^{-1}hr^{-1}}$\\$i_{r}:0$\\$\rm{q_{cic}:  0.0013\ e^{-}pix^{-1}frame^{-1}}$ \\($t_{\rm{frm}}=100s)$ }   &  \makecell[l]{$\rm{i_{d}:0}$\\  $\rm{i_{r}: 0.5\ e^{-}pix^{-1}frame^{-1}}$ \\($t_{\rm{frm}}=1000s)$ \\ $\rm{q_{cic}:0}$ } \\
\enddata
\tablenotetext{1}{Astrophysical parameters of targets are listed in table~\ref{tab:rvs} }
\tablenotetext{2}{$\tau_{core}$ is the core throughput of coronagraph or starshade, not including optical throughput due to other optical opponents and detector quantum efficiency. }
\tablenotetext{3}{$\tau_{\rm{optical}}$ is the throughput of optical component of the system.}
\tablenotetext{4}{$n_{\rm{pix}}$ is the number pf pixels contributing to one spectral element in IFS simulation.}
\tablenotetext{5}{$i_{d}$: dark current count rate, $i_{c}$: readout noise count rate, $q_{cic}$: clock induced charge, $\rm{t_{frm}}$: integration time per frame.}
\tablenotetext{6}{Other parameters: coronagraph contrast $10^{-10}$; Nyquist wavelength: VIS 540 nm, NIR 1000 nm; Telescope diameter: 6m }
\end{deluxetable*}

\section{Simulation Setup}\label{sec:instru}

\subsection{General Simulation Setup}

Table~\ref{tab:rvs} lists the astrophysical properties of our fiducial target system. We consider the target as an Earth-like planet around a sun-like star (4.83 mag G0V star) at 1AU at 10 pc away. 

As shown by the models presented in Appendix~\ref{sec:plab}, the signal-to-noise ratio is determined by instrument parameters including the telescope size, the detector noise rate, the throughput of the coronagraph.  To understand their effect on the relative sensitivity of an IFS to an iFTS to detecting an Earth-like planet around a Sun-like star, we first simulate a general space-based mission with varying instrument parameters including coronagraph throughput $\tau_{core}$ and the pixel number assigned to each IFS spectral element $n_{pix}$. Simulations $ 2\& 3$ in Table~\ref{tab:so}  present two such examples in a visible band (720-800 nm) and an NIR band (1050-1200 nm), respectively. We consider $10\%$ bandwidth as it becomes increasingly challenging for the wavefront control system to maintain high-contrast across broader bandpasses. We assume a telescope diameter of 6 meter and coronagraph contrast of $C=10^{-10}$. We consider the system throughput as the product of optical throughput $\tau_{optical}$, coronagraph core throughput $\tau_{core}$  and detector quantum efficiency $\tau_{\rm{QE}}$ (details see Appendix~\ref{sec:ST}). We use the average optical throughput of $\tau_{optical}=23.8\%$ at visible wavelengths and $\tau_{optical}=32.4\%$ in the NIR wavelengths from the end-to-end studies of LUVOIR \citep{LUVOIRTeam}. We also adopt the spectral resolving power values from LUVOIR \citep{LUVOIRTeam}, which are 140 in the optical and 70 in the infrared. The coronagraph core throughput ranges between $3\%$ and $40\%$, while the IFS $n_{pix}$ varies from 2 to 10. 

 
Table~\ref{tab:so} also presents the detector noise rate in the simulations. We consider two levels of detector noise, referred to as ``current'' and ``future'' levels. The current level refers to state of the art (SoA) values from detectors used in operating or soon-to-launch space missions. For the future detector noise, we adopt HabEx and LUVOIR requirements. In the visible bands (500-1000nm), we consider the dark current and clock-induced charge (CIC) of the electron multiplying charge-coupled device (EMCCD). For the current level, we adopt values from the coronagraph instrument on Roman, which uses the e2v CCD201-20 EMCCD for both imaging and spectrograph cameras. There are several reference values of its dark current\footnote{Recently, it was found that operating the EMCCD in inverted mode led to a remarkable decrease in dark current to   $0.05-0.1\ \rm{e^{-}\ hr^{-1}\ pix^{-1}}$. (P. Morrissey, private communication).  This may become the flight baseline.  However, our detector numbers are based on the most recent published work, \citet{morrissey2023flight}.  If the dark current remains this low, CIC will be the dominant contributing noise, but there will be no change to the conclusions of this paper regarding the unsuitability for an iFTS in the optical.} \citep{morrissey2023flight}: current best estimate (CBE, $0.77\ \rm{e^{-}\ hr^{-1}\ pix^{-1}}$), beginning of life requirement  ($\rm{BOL\_REQ}$, $1.65\ \rm{e^{-}\ hr^{-1}\ pix^{-1}}$) and end of life requirement ($\rm{EOL\_REQ}$, $3\ \rm{e^{-}\ hr^{-1}\ pix^{-1}}$). Similarly, the CIC at beginning of life is 0.01 e-/pixel/frame. In this case, the dark current is the dominant noise source as it is between 0.77-3 electron/pix/hr, whereas CIC is only 0.36 electron/pix/hr assuming 100s exposure time per frame. In addition, we adopt a constant CIC as the photon counts per pixel per frame from a Earth like planet and the backgrounds are $\ll 1 \rm{e^{-}\ \rm{frame}^{-1}\ pix^{-1}}$ for both an IFS and an iFTS. For the future level, we adopt the dark current values required by LUVIOR and HabEx as $0.1\ \rm{e^{-}\ \rm{frame}^{-1}\ pix^{-1}}$ \citep{Gaudi2020,LUVOIRTeam}.

In the near-IR  (1000-2000 nm), we consider the HAWAII H2RG as the state-of-the-art. The total effective noise of H2RG detectors on JWST is  $5.9\ \rm{e^{\text{-}}\ frame^{\text{-}1}\ pix^{\text{-}1}}$ with an integration time per frame of 1000s and $10\ \rm{e^{\text{-}}\ frame^{\text{-}1}\ pix^{\text{-}1}}$ with an integration time per frame of 100s \citep{JWSTnoise2022}. Because readout noise is the dominant noise source of NIR detector, we use a readout noise of $5.9\ \rm{e^{\text{-}}\ frame^{\text{-}1}\ pix^{\text{-}1}}$ with an integration time per frame of 1000s and ignore the dark current. HabEx baselined (and LUVOIR listed as backup) HgCdTe linear mode avalanche photodiode array detectors (LMAPD)for the NIR channels.  \footnote{Both the HabEx and LUVOIR report appear to suggest that current LMAPDs like the small-format SAPHIRA arrays can achieve dark currents of 1.5e-3 electrons per pixel per second simultaneously with sub-electron read noise, and thus development of larger arrays is all that is needed.  This is false.  When the effective read noise is reduced to sub-electron levels through avalanching, the dark current increases by approximately three orders of magnitude.  There are no  space-qualified infrared arrays that can simultaneously achieve low dark current and read noise\citep{Atkinson2017,Goebel2018}, though there are development efforts underway \citep{claveau2022first}}
HabEx required $<0.002 \rm{e^{\text{-}}\ s^{\text{-}1}\ pix^{\text{-}1}}$ dark current and $\ll 1\ \rm{e^{\text{-}}\ \rm{frame}^{\text{-}1}\ pix^{\text{-}1}}$ read noise. LUVOIR required  $3\ \rm{e^{\text{-}}\ frame^{\text{-}1}\ pix^{\text{-}1}}$ readout noise with $0.001 \rm{e^{\text{-}}\ hr^{\text{-}1}\ pix^{\text{-}1}}$. Here, we adopt a readout noise of $0.5\ \rm{e^{\text{-}}\ frame^{\text{-}1}\ pix^{\text{-}1}}$ as the future level NIR detector noise.

\subsection{Simulation setup for \text{HWO} in various bands}
Simulation $4a\&$b in Table~\ref{tab:so} present the instrument parameters of simulations for a 6-meter direct imaging telescope. We use the baseline designs and requirements from LUVOIR and HabEx \citep{LUVOIRTeam, Gaudi2020} as references for other parameters, with a coronagraph throughput of $10\%$. We use filter specifications and optical throughput of LUVOIR because the optical design is incomplete for the Habitable Worlds Observatory.

\section{Results}\label{sec:resu}

\subsection{General simulation results in Visible}\label{sec:5_1}
\begin{figure*}
    \centering
    \includegraphics[width=\linewidth]{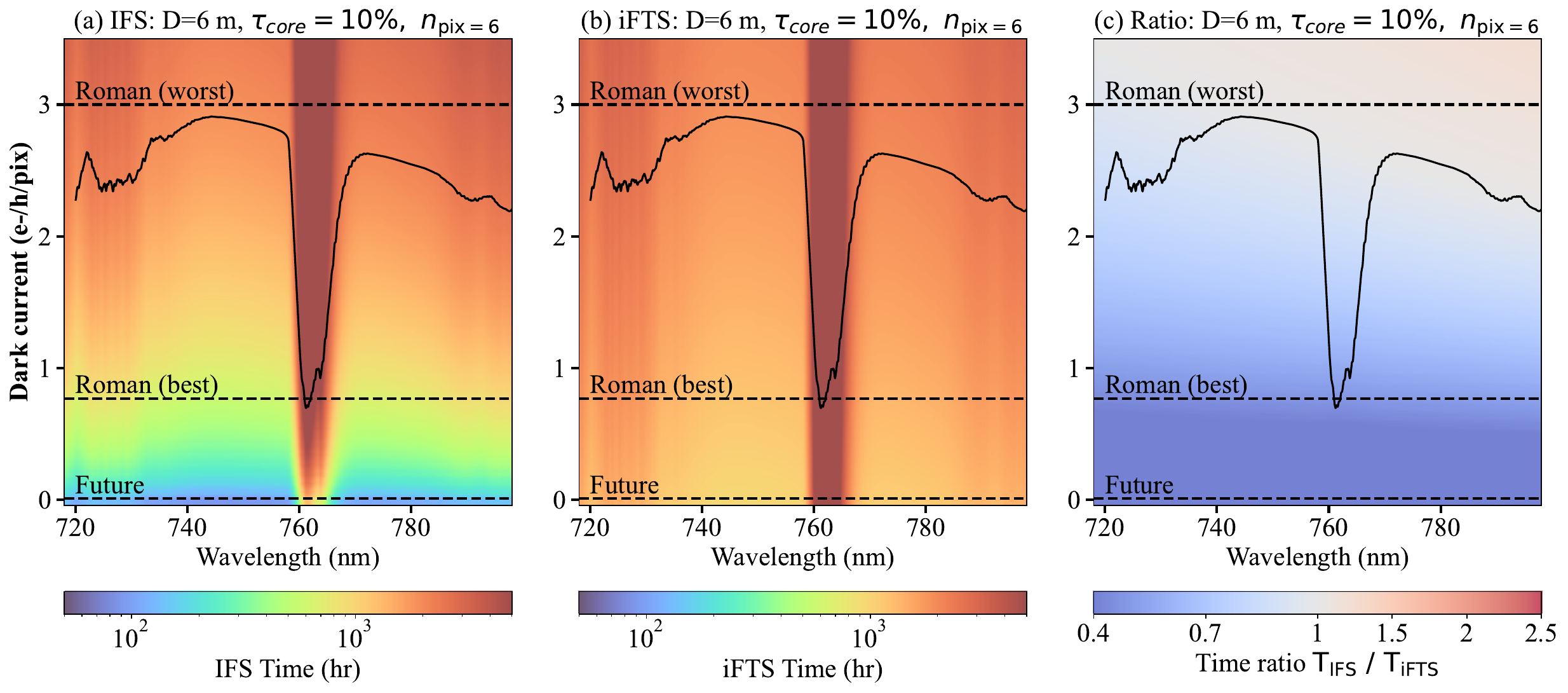}
    \caption{(a): the required exposure time to achieve an SNR of 5 using an IFS as the function of EMCCD dark current in a visible band (720-800 nm, R=140); (b) the same as (a), but for an iFTS; (c) the required exposure time ratio between an IFS and an iFTS $\rm{T_{IFS}/{T_{iFTS}}}$ as the function of EMCCD dark current. We use fixed instrument parameters including diameter$D=6 m$, coronagraph throughput $\tau=10\%$, and IFS $n_{pix}=6$. The target is an Earth-twin at 10 pc way. The normalized planet spectrum is superimposed on the three panels. The horizontal dashed lines correspond to several reference values for detector noise, including the state-of-art dark current from EMCCD on Roman space telescope (best and worst scenarios) and detector noise requirements of future missions.  }
    \label{fig:figure6}
\end{figure*}
\begin{figure*}
    \centering
    \includegraphics[width=\linewidth]{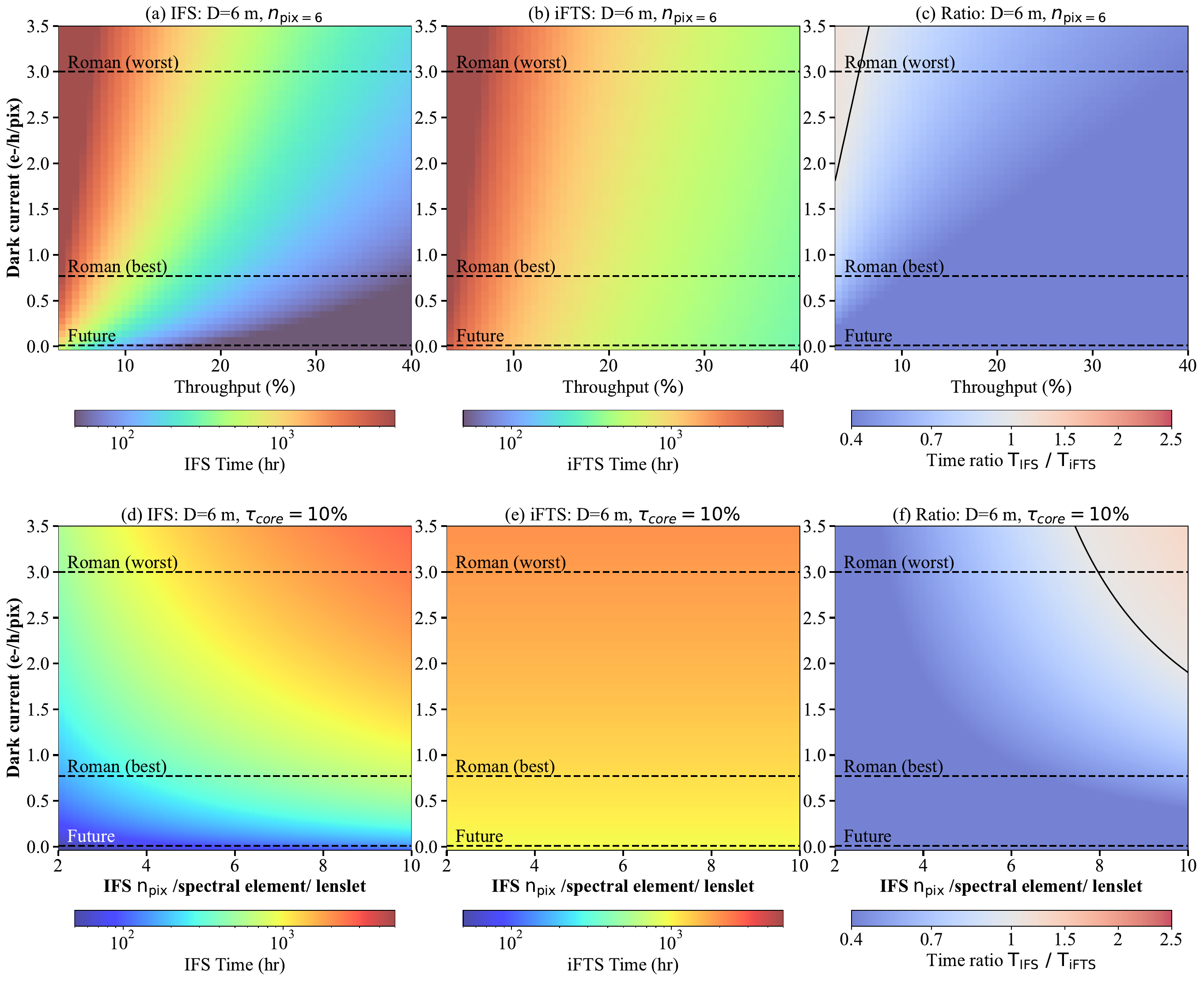}
    \caption{For the same band in figure~\ref{fig:figure6}, we further compare the IFS and iFTS required time at the continuum ($\lambda = 750\ nm$) with various instrument parameters. (a)-(c): the required exposure time to achieve an SNR of 5 using an IFS, an iFTS and the time ratio between an IFS and iFTS as the functions of EMCCD dark current and coronagraph throughput. (d)-(f): the required exposure time to achieve an SNR of 5 using an IFS, an iFTS and the time ratio between an IFS and iFTS as the functions of EMCCD dark current and IFS detector sampling number $n_{IFS}$.  Solid lines mark where the time ratio equals to one. The horizontal lines correspond to several reference values for detector noise, including the state-of-art dark current from EMCCD on Roman space telescope (best and worst scenarios) and future detector noise required by future missions.}
    \label{fig:figure6_2}
\end{figure*}

\begin{figure*}
    \centering
    \includegraphics[width=\linewidth]{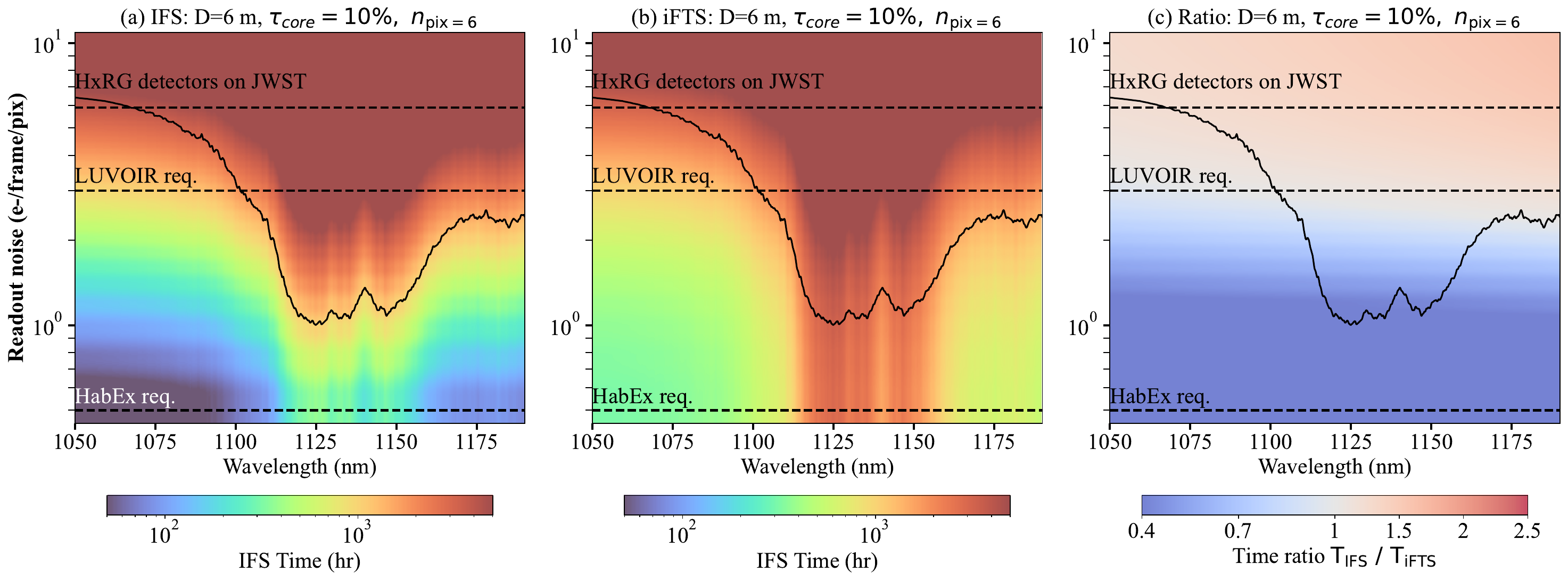}
    \caption{(a): the required exposure time to achieve an SNR of 5 using an IFS as the function of readout noise in a NIR band (1050-1200 nm, R=70); (b) the same as (a), but for an iFTS; (c) the required exposure time ratio between an IFS and an iFTS $\rm{T_{IFS}/{T_{iFTS}}}$ as the function of readout noise. We use fixed instrument parameters including diameter$D=6 m$, coronagraph throughput $\tau=10\%$, and IFS $n_{\text{pix}}= 4$. The target is an Earth-twin at 10 pc way. The normalized planet spectrum is superimposed on the three panels. The horizontal dashed lines correspond to several reference values for detector noise, including the state-of-art dark current from HxRG on JWST and future detector noise required by future missions.}
    \label{fig:figure6_3}
\end{figure*}
\begin{figure*}
    \centering
    \includegraphics[width=\linewidth]{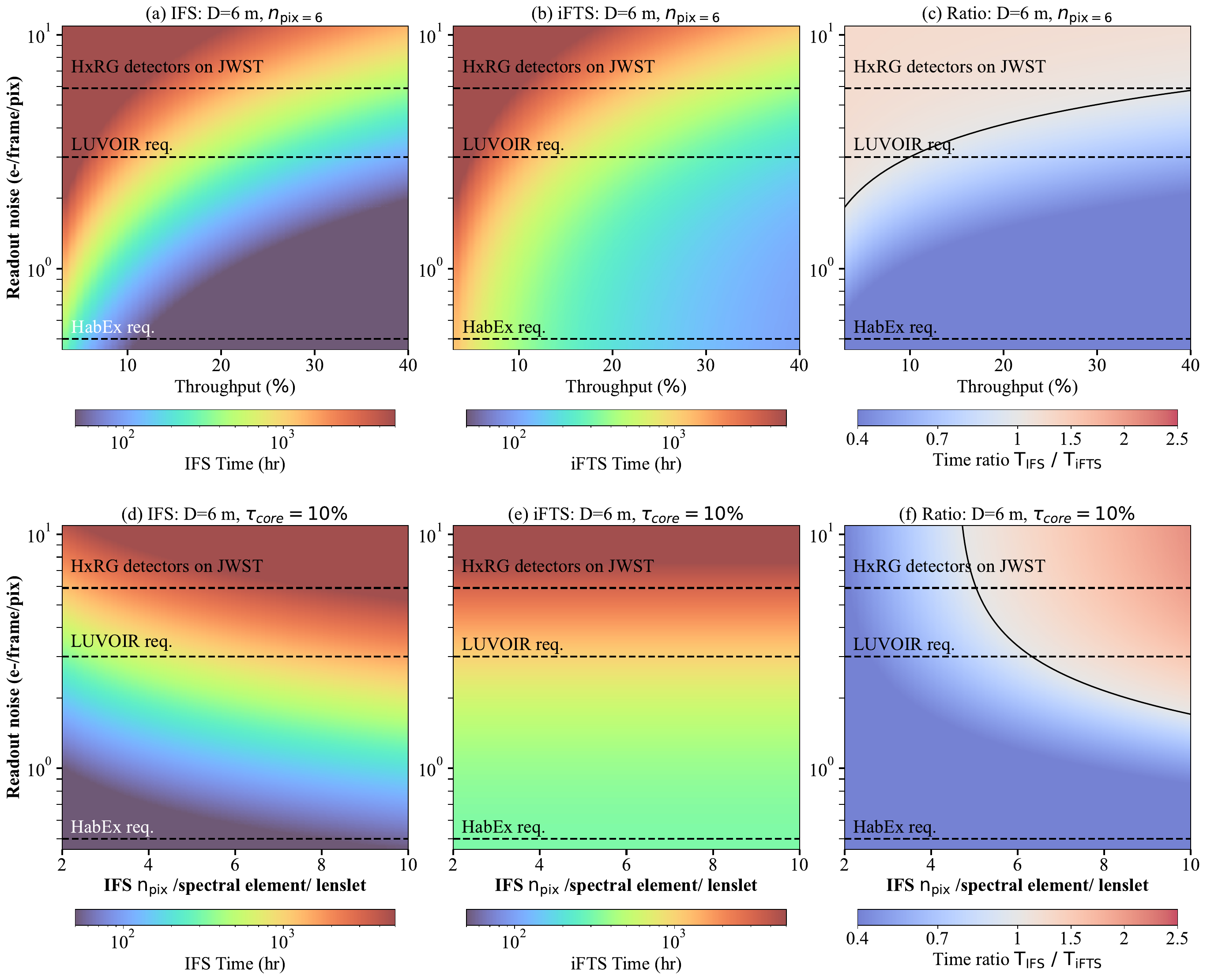}
    \caption{For the same band in figure~\ref{fig:figure6_3}, we further compare the IFS and iFTS required time at the continuum ($\lambda = 1050\ nm$) with various instrument parameters. (a)-(c): the required exposure time to achieve an SNR of 5 using an IFS, an iFTS and the time ratio between an IFS and iFTS as the functions of readout noise and coronagraph throughput. (d)-(f): the required exposure time to achieve an SNR of 5 using an IFS, an iFTS and the time ratio between an IFS and iFTS as the functions of readout noise and IFS detector sampling number $n_{\rm{IFS}}$.  Solid lines mark where the time ratio equals to one. The horizontal dashed lines correspond to several reference values for detector noise, including the state-of-art dark current from HxRG on JWST and future detector noise required by future missions.}
    \label{fig:figure6_4}
\end{figure*}

Figure~\ref{fig:figure6} (a) $\&$ (b) display the required exposure times of an IFS and an iFTS to achieve the SNR of 5 as the functions of detector noise and wavelength in a visible bandpass. We vary the dark current rate for the visible camera from state of the art value (EMCCD on \textit{Roman-CGI}) to the future value required by LUVOIR/HabEx (detector noise details see Table ~\ref{tab:so}). We focus on the band from 720 nm-800 nm due to the strong $\rm{O_{2}}$ line at 760 nm. Both the IFS and iFTS need significantly longer exposure times in the absorption feature compared to the continuum. For the continuum, the IFS exposure time exhibits a strong dependency on the dark current, decreasing from $\ttilde 2000$ hours with \textit{Roman-CGI}'s worst detector noise ($3 \ \rm{e^{\text{-}}\ hr^{\text{-}1}\ pix^{\text{-}1}}$) to $\ttilde 500$ hours with Roman's best detector noise ($0.77\ \rm{e^{\text{-}}\ hr^{\text{-}1}\ pix^{\text{-}1}}$). If the detector noise is further reduced to $0.1\ \rm{e^{\text{-}}\ hr^{\text{-}1}\ pix^{\text{-}1}}$, as required by LUVIOR and HabEx, the IFS exposure time at the continuum will be $ \ttilde 100$ hours.

In contrast, the iFTS exhibits lower sensitivity to the dark current, requiring $ > 1000$ hours of exposure time across varying dark current values. As elaborated in section~\ref{sec:CMW}, the iFTS enjoys the benefit of reduced sensitivity to detector noise, yet it is more susceptible to the effects of photon noise. In optical wavelengths, the impact of photon noise outweighs the advantages of reduced detector noise for the iFTS, limiting its performance. Figure~\ref{fig:figure6} (c) shows the required exposure time ratio between an IFS and an iFTS in the same bandpass as Figure~\ref{fig:figure6} (a)$\&$(b). It is evident that an IFS outperforms an iFTS in most cases. The time ratio exhibits a weak dependence on the wavelength because the IFS detector noise increases at longer wavelengths.

Figure~\ref{fig:figure6_2} (a)-(c) compare the performance of an IFS and an iFTS with varied detector noise and coronagraph throughput for a 6 meter telescope.  The simulation is conducted within the same bandpass as depicted in Figure~\ref{fig:figure6}, and the exposure time considered is that of the continuum. For reference, \textit{Roman-CGI} has a low coronagraph throughput of $\ttilde3\%$, whereas studies on coronagraphs for future missions are underway to achieve higher throughput. When the coronagraph throughput $\tau_{core}$ is below $5\%$, both the IFS and iFTS required exposure times are over 5000 hours, rendering the detection unfeasible. As the coronagraph throughput increases, the IFS and iFTS performance both improve. Nonetheless, in cases of higher throughput, the iFTS experiences a higher penalty from photon noise compared to the IFS. This is evident from Figure~\ref{fig:figure6_2}(c), which illustrates that the IFS would surpass the iFTS in performance when the throughput exceeds around $10\%$.
\begin{figure*}
    \centering
    \includegraphics[width=\linewidth]{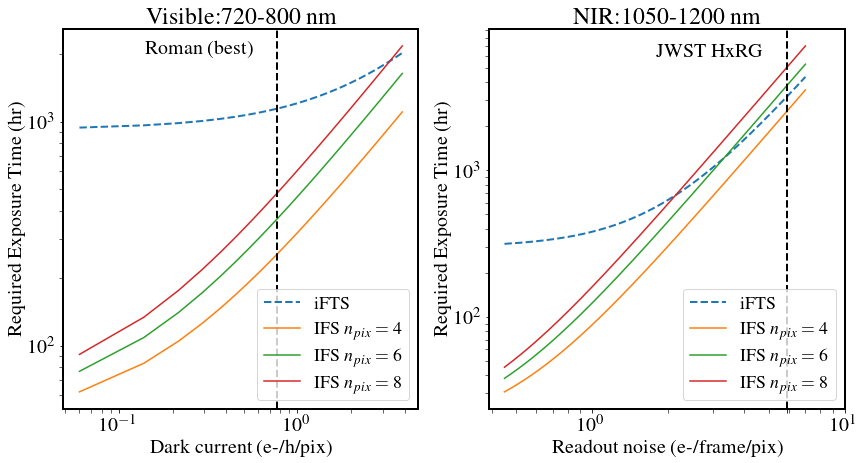}
    \caption{The required exposure time to achieve a SNR of 5 as the function of the detector noise for an iFTS (dashed line), solid lines are for an IFS with varied $n_{pix}$ (solid lines). The left is for the visible band (720-800 nm) same as figure~\ref{fig:figure6} and figure~\ref{fig:figure6_2}, the required time is calculated at the continuum. The right is for the NIR band (1050-1200 nm) same as figure~\ref{fig:figure6_3} and figure~\ref{fig:figure6_4}, the required time is calculated at the continuum. The dashed lines correspond to the state-of-art detector noise from EMCCD on Roman-CGI (best scenario) and  the readout noise for HxRH detector on JWST.   }
    \label{fig:figurea5}
\end{figure*}

Figure~\ref{fig:figure6_2} (d)-(f) investigate how the IFS detector sampling influences the relative performance of an iFTS and an IFS for a 6 meter telescope. As an IFS retrieves spectral information through dispersing photons into different pixels, the total detector noise depends on the number of pixels per spectral element per lenslet  $n_{pix}$. As shown in section~\ref{sec:AF}, $n_{pix}$ is the product of pixel numbers in vertical and horizontal directions. HabEx and LUVOIR both use 6 pixels ($2\times 3$) and 4  pixels ($2\times 2$) per spectral element per lenslet in the optical and NIR bandpasses for coronagraph mode, respectively \citep{LUVOIRTeam, Gaudi2020}. We varied $n_{pix}$ from 2 to 10 in our simulations. In the visible bandpass, an IFS always outperforms an iFTS if the IFS uses less than 6 pixels per spectral elements per lenslet. But the IFS performance degrades as  $n_{pix}$ increases, being less effective than an iFTS if $n_{pix}>6$ and dark current rate $> 2\ \rm{e^{\text{-}}\ hr^{\text{-}1}\ pix^{\text{-}1}} $. 



In summary, these simulations reveal that an iFTS is less effective than an IFS at optical wavelengths for a 6 meter telescope, because an iFTS is fundamentally limited by photon noise.

\begin{figure*}
    \centering
    \includegraphics[width=0.9\linewidth]{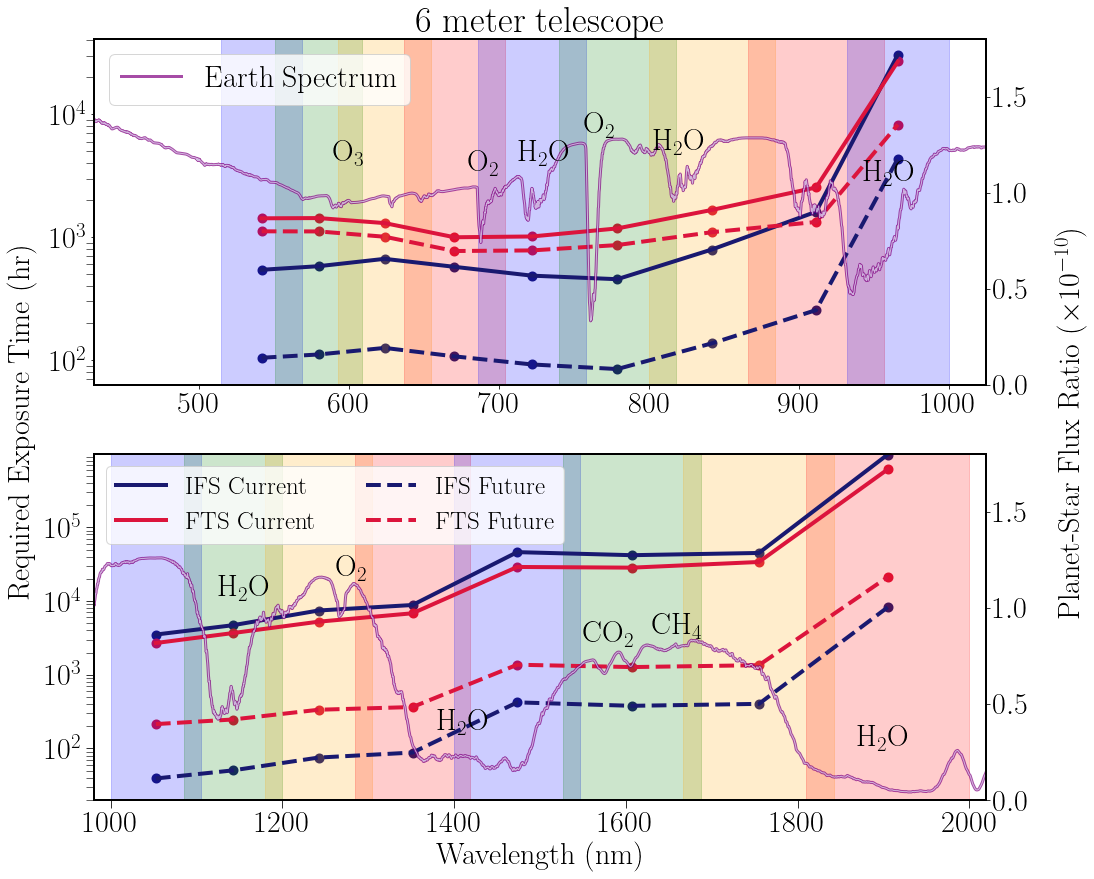}
    \caption{The required integration time to achieve $\rm{S/N=5}$ for a 6 m telescope equipped with an iFTS (red) and an IFS (dark blue) at different spectroscopy bands in the optical (top panel) and NIR (bottom panel). The filter bands are shown as colorful shaded area and  every point corresponds to the central wavelength of every band. In the visible wavelengths, we use Roman current best estimate dark current (solid lines) and the required dark current by future mission (dashed lines). In the NIR, there are also two types of readout noise: JWST current best estimate readout noise ( solid lines) and the future level of readout noise (dashed lines). The target is a 1 $R_{\oplus}$ planet orbiting  \textbf{a 4.83 mag G0V star} at 1 AU from 10 pc away. Instrument parameters are listed in 3 (a) and (b) in Table~\ref{tab:so}}
    \label{fig:figure9}
\end{figure*}
 
\subsection{General simulation results in NIR}

Figure~\ref{fig:figure6_3} compares the require exposure times of an IFS and an iFTS in a NIR band containing a $\rm{H_{2}O}$ absorption feature from 1050 nm to 1200 nm. For the NIR detectors, the readout noise is the dominant noise source for which we vary from the state of the art HxRG detectors on JWST to future LMAPD-like level, in both cases with a frame time of 1000 s  (detector noise details see Table ~\ref{tab:so}). LUVOIR and HabEx reports use an $n_{\rm{pix}}=4$ design for the IFS in the NIR. As detailed in section~\ref{sec:CMW}, when $n_{\rm{pix}}=4$, an iFTS will have equivalent detector noise to an IFS but higher photon noise. Consequently, an iFTS is always less effective than an IFS with $n_{\rm{pix}}=4$.  Hence, in our simulations, we employ an IFS $n_{\rm{pix}}=6$  to investigate the relative sensitivity within this setup.

Figure~\ref{fig:figure6_3} reveals that the exposure times needed for the IFS and iFTS in the NIR band are longer than the visible band in Figure~\ref{fig:figure6}. This is because NIR detectors have higher detector noise, primary the readout noise. Specifically, at the continuum of the $\rm{H_{2}O}$ absorption band, if the readout noise is above  $3\ \rm{e^{\text{-}}\ frame^{\text{-}1}\ pix^{\text{-}1}}$, both the IFS and iFTS demand exposure times over 1000 hours, with the iFTS displaying slightly better efficiency. However, as the readout noise decreases to levels below around $3\ \rm{e^{\text{-}}\ frame^{\text{-}1}\ pix^{\text{-}1}}$, the IFS surpasses the iFTS in performance, requiring only half the exposure time of the iFTS.

Figure~\ref{fig:figure6_4} is similar to Figure~\ref{fig:figure6_2}, but specifically for the NIR band ranging from 1050 nm to 1200 nm. It is evident that both IFS and iFTS benefit from increased coronagraph throughput and reduced readout noise, resulting in decreased required time. Considering the readout noise of the HxRG detector on JWST,  the time required by IFS and iFTS will diminish to less than 1000 hours when throughput surpasses $\ttilde25\%$.In these scenarios, iFTS demonstrates $\ttilde 20\%$ better efficiency. Should the readout noise adhere to the specifications outlined in the LUVOIR report ($3\ \rm{e^{\text{-}}\ frame^{\text{-}1}\ pix^{\text{-}1}}$), an iFTS requires less time if the coronagraph throughput is below $\ttilde 10\%$. However, the opposite occurs when the throughput is larger than $\ttilde 10\%$. As the readout noise is further reduced below 1 e-/pix/frame an IFS emerges as the more efficient choice.
 
Figure~\ref{fig:figure6_4} (d)-(f) shows that the performance comparison between an iFTS and an IFS also depend on the IFS detector sampling design. Consistent with the argument in section~\ref{sec:CMW}, if an IFS is assigned  $\leq 4$ pixels per spectral element per lenslet, it always wins over an iFTS. (Note that the plot limit is set at 2 pixels per spectral element per lenslet, which corresponds to a lenslet being imaged to a single row of pixels on the detector with spectral sampling at the Nyquist limit.)  However, if the IFS  $n_{pix}> 4$, the IFS performance will deteriorate. When IFS  $n_{pix}> 4$ and readout noise is above $2\ \rm{e^{\text{-}}\ frame^{\text{-}1}\ pix^{\text{-}1}}$, an iFTS would be more efficient by $\ttilde30\%$.

In summary, the comparison of the iFTS and IFS performance in the near-IR channel depends on the instrument parameters, especially the number of pixels per spectral element per lenslet used by the IFS. Furthermore, we have implemented a frame integration time of 1000 seconds for NIR bandpass in our simulations. A longer integration period helps to reduce readout noise, while resulting in a higher number of cosmic ray hits.  While a comprehensive study to determine the optimal integration time per frame is necessary, this task falls outside the scope of this paper.

\subsection{Exposure times with decreased detector noise}

Figure~\ref{fig:figurea5} presents the required exposure time  to achieve a SNR of 5 at continuum by an iFTS and an IFS as the function of detector noise. As discussed in section~\ref{sec:5_1}, an iFTS is limited by the photo noise in the visible, always requiring $>1000$ hours exposure time. For an IFS, to reduce the exposure time to the order of $100$ hours, the dark current needs to be reduced by a factor of 5 from the current published level \citep{morrissey2023flight}. In the NIR, if the readout noise remains $>2-3\ \rm{e^{\text{-}}\ frame^{\text{-}1}\ pix^{\text{-}1}}$ and  IFS $n_{pix}>4$, an iFTS will use at least $\ttilde20\%$ time although the absolute time scale is above 1000 hrs for the two spectrographs in the case. The required exposure time by an IFS can be decreased to the order of $100$ hours if readout noise is reduce to below $1\ \rm{e^{\text{-}}\ frame^{\text{-}1}\ pix^{\text{-}1}}$ with a frame time of 1000s. 

\subsection{Results for HWO in various bands }
Figure~\ref{fig:figure9} presents the required exposure time for an IFS and an iFTS to achieve a SNR of 5 in different filters in the optical and NIR. We compare the required time with current and future level detector noise, respectively. In the optical, the required integration time rises sharply at channels longer than 800 nm, because the detector quantum efficiency drops significantly at the red wavelengths (see Appendix~\ref{sec:Dqe}). In addition, the detector noise rises as the wavelength increases because PSF size becomes larger and involves more detector pixels.  Thus, we mainly focus on the channels shorter than 800 nm. In the optical range, photon noise is the dominant source of noise in an iFTS, so reducing detector noise does not significantly improve performance. The iFTS required exposure time is $\ttilde1100 $ hrs in channels shorter than 820 nm with current dark currents, and $\ttilde1000$ hrs with future dark currents. In contrast, an IFS currently requires approximately $\ttilde700 $ hrs for operation, but a substantial reduction in dark current greatly enhances its performance, reducing the required time to $\ttilde100 $ hrs with a future level of detector noise. The most promising band in the visible will be the channel around 760 nm with a oxygen line.

In the NIR, the required exposure time is longer at bands with wide absorption features, mostly from $\rm{H_{2}O}$. As the wavelength increases, the PSF aperture also expands, prompting the involvement of a greater number of pixels in both the IFS and iFTS methods. This augmented pixel utilization contributes to higher detector noise, consequently increasing the required exposure time. The most promising band for detection is from 1050nm to 1200nm  with a $\rm{H_{2}O}$ absorption feature. 

With an IFS $n_{pix}$ of 6, an IFS requires longer exposure times than an iFTS to achieve a SNR of 5 with current-level readout noise. However, the absolute time scale is beyond a few thousand hours and even higher than that in the longer wavelengths, making the detection unfeasible. If the readout noise is reduced to the $0.5\ \rm{e^{\text{-}}\ frame^{\text{-}1}\ pix^{\text{-}1}}\ (t_{\rm{frm}}=1000\ s)$, an IFS will be more efficient. At the band from 1050-1200nm with a $\rm{H_{2}O}$ absorption line, an IFS will require $ <100$ hrs to achieve a SNR of 5. The required time increases at longer wavelengths. It takes at least a few hundred hours in the bands from 1300-1500 nm and even a few thousand hours in he bands from 1800 nm to 2000nm, both of which encompass broad absorption features attributed to $\rm{H_{2}O}$. Note that the exposure times needed for the bluest filters in near-infrared wavelengths are shorter than those for the reddest filters in the visible wavelengths. This is due to the decline in EMCCD quantum efficiency, which falls to approximately $30\%$ beyond 900 nm, and the presence of absorption features. Additionally, the resolving power used in the optical range is 140, which is higher than the 70 resolving power used in the NIR range.

\section{Discussion}\label{sec:diss}

\subsection{Practical operational concerns}
Our results have been derived from a purely ``point-source, photon budget'' perspective, keeping as many instrument parameters between the two spectrograph architectures as equal as possible.  For example, iFTS advantages in throughput (reasonably estimated at $\sim$15\% higher, \citep{bennett1999critical}) were reduced to IFS values, and iFTS-specific issues like field-dependent modulation efficiency, beamsplitter chromaticity, polarization effects, etc were ignored.  We have also not discussed challenges and opportunities with operating an iFTS as part of a space-based high-contrast imaging system. Here we briefly discuss such practical considerations, focusing on wavefront control, sensitivity to space radiation effects, and the flexibility offered by the instrument.

To properly understand the feasibility of integrating an iFTS with an exoplanet imaging system, simulated two-dimensional images of a coronagraph,  wavefront control system, and iFTS would be needed, which are beyond the scope of this work. This paper is a first step in exploring the potential use of an iFTS for direct imaging exoplanets, with an analysis based on point photometry and the assumption that an iFTS works with upstream instruments. Future work is to add two-dimensional simulations, including realistic tolerances, of how electric fields propagate through coronagraph and wavefront control systems before they are fed into the spectrograph.

\subsubsection{Wavefront Control}
Operation of a coronagraph in space requires nanometer to picometer level control of wavefront error.  To achieve this, space coronagraphs use the science camera in conjunction with the deformable mirrors to iteratively sense and correct the electric field at the detector to achieve a zone of very high contrast, known as ``digging a dark hole''.  Once the dark hole is established, it needs to remain stable.

The situation is more complicated with an iFTS due to the moving mirror.  While a dark hole could be established at the zero-path difference of the interferometer, an open question is whether the contrast would be stable during a mirror scan.  Rough arguments suggest that it would.  First, the main optical errors corrected by the dark hole are upstream of the coronagraph, with downstream optics playing a secondary role.  Second, the tolerances already demonstrated by iFTS instruments on  mirror stability and position are close to those required for high contrast imaging--for example, SITELLE achieves sub-nm accuracy over a 1 cm scan using a 15 cm flat mirror \citep{grandmont2012final}. Finally, phase-to-amplitude wave mixing happens on distances comparable to the Talbot length, which is \ttilde meters in a high contrast imaging system, much longer than any mirror scan distance \citep{mazoyer2017active}. This suggests the dark hole contrast should not be meaningfully affected.

The need for wavefront control also negates one of the principal advantages of an iFTS, which is its broad bandwidth, limited by reflectance properties of the mirrors and detector quantum efficiency.  However, wavefront control algorithms struggle with broadband light, and are not expected to operate with more than 10-20\% bandwidth \citep{LUVOIRTeam, Gaudi2020}.  Thus upstream filters must limit the bandwidth through the iFTS.

\subsubsection{Detector Effects}
Space radiation effects on detectors can play a dominant role in instrument performance.  We briefly comment on the effects of cosmic rays and accumulated radiation damage on iFTS performance.

The focal-plane image in an iFTS is orders of magnitude smaller and brighter than in an IFS since an iFTS does not disperse photons.  The chance of a cosmic ray hit on the planet signal is thus proportionally reduced, though suffers a minor penalty from the two detectors.  On the other hand, while a cosmic ray in an IFS only affects the the section of the 2-d spectrum it hits, in an iFTS, a cosmic ray affects the entire spectrum, due to the effect of Fourier transforming the interferogram.  Special care would need to be taken to clean the interferogram of cosmic rays before extracting the spectrum. 

Accumulated damage from cosmic rays causes degradation in detector parameters like dark current.  For example, the dark current in the least noisy detector of the wide-field planetary camera of the Hubble Space Telescope quadrupled between 1995 and 1998, with rates of increase closely agreeing with independent measurements of galactic and solar cosmic rays \citep{mcmaster2008wide}.  Such increases in dark current would substantially affect IFS sensitivity, while an iFTS would only have a modest decrease in performance, as it is mainly limited by photon noise (see Figures \ref{fig:figure6_2}).

\subsubsection{Variable resolving power of iFTS}
There is additional flexibility in the iFTS for spectroscopic exoplanet surveys beyond that afforded by a camera and IFS.  

First, at the zero-path difference of the instrument, the iFTS functions as a conventional camera, with all the light going to one detector, with minimal efficiency loss due to the extra surfaces and no noise penalty over a conventional camera.  Even when scanning, broadband imaging data is simultaneously available by adding the two output port images, though this increases the detector noise in the imaging data by $\sqrt{2}$.  This allows for dynamic planet identification and localization during long observations.  The increase in image brightness in an iFTS compared to an IFS is advantageous for this task.  The simultaneous imaging and spectroscopy allows for simultaneous planet search and characterization, which is qualitatively different than search strategies with a camera and IFS.  However, the time advantage conferred may be minor as these missions are unlikely to rely on blind searches.

The flexible resolving power is also potentially advantageous.  As the mirror scan distance increases, the resolving power increases, allowing for ``dynamical'' vetting of interesting sources at low resolving power.  The resolving power may be adjusted to observe Earth-like planets with low resolving power mode and bright young giant plants with medium/high resolving power mode.  Similarly, if it is found that more resolving power is a science need on a particular target, the mirror scan may simply be restarted from the last position.

The previous two advantages suggest a high-level analysis of observations with an iFTS and their effects on survey yield (eg, \citet{stark2014maximizing, savransky2017exosims}), are needed to understand how and if this flexibility can be leveraged.

\section{Conclusion}\label{sec:concs}
In this work, we have compared the performance of an imaging Fourier transform spectrograph to an integral field spectrograph for the challenging task of obtaining spectra from an Earth-like planet. To do this, we created analytical and numerical models to simulate the iFTS and IFS observations of exoplanet reflective spectra from visible to near-infrared wavelengths using an future space direct imaging telescopes with varied instrument parameters. While the relative performance of the two instrument concepts depends on telescope, instrument, detector, and planetary system parameters, several general conclusions can be drawn.

First, in the optical, an IFS outperforms an iFTS to characterize Earth-like planets around nearby stars, since the latter is limited by photon noise.  Reductions in dark current by a factor of 5 from current published values \citep{morrissey2023flight} will improve the performance of an IFS, allowing it to obtain the spectra from Earth-like planets within a reasonable time (below $\ttilde 100$ hrs) in the bluest wavelengths, whereas an iFTS has a more minor gain in performance.  to the decreasing of EMCCD quantum efficiency and larger PSF size with the increase of wavelength, the most promising wavelengths are below 800 nm, especially around 760 nm with an $\rm{O_{2}}$ line. The most promising wavelengths in NIR for HWO coronagraphy from an integration time point of view are from 1050-1200 nm with a water asorption feature.

On the other hand, in the near-IR, the relative sensitivity between an IFS and an iFTS critically  depends on the instrument design. An IFS will always be more efficient if it uses 4 or less pixels per spectral element per lenslet. If using more than 4 pixels per spectral element per lenslet, the IFS performance is less efficient by at least $\ttilde 20 \%$ than an iFTS when the readout noise is higher than $2-3\ e^{-}\ pix^{-1}\ \rm{frame}^{-1}\ (\rm{t_{frm}=1000s})$, though the \textit{absolute} sensitivity depends strongly on the readout noise level. Therefore, with currently available HxRG NIR detctors, an iFTS woule be more efficient than an IFS. But if the effective readout noise levels is below 2 electrons, the required time by an IFS becomes much shorter than an iFTS,reduced to 200-1000 hrs at the shortest near-IR bandpass (1050-1200 nm). The readout noise needs to be reduced to sub-electron level ($<1\ e^{-}\ pix^{-1}\ \rm{frame}^{-1}$) to achieve a SNR of 5 within 100 hrs. The required exposure time rises with the increase of wavelength as detector noise is higher with larger PSF.


Given the high priority of a space high contrast imaging mission, it is encouraging to see that at least one other instrument concept exists that can provide comparable sensitivity to an IFS at infrared wavelengths, where most of the deep biosignature features exist.  Other aspects of the iFTS, like simultaneous imaging and spectroscopy, flexible resolving power, and the ability to use orders-of-magnitude smaller-format detectors, may present new opportunities for such a mission, though these features were not explored in this work.  We hope the results of this paper motivate consideration of an iFTS in future design trades and yield simulations of this important future mission.





\section*{Acknowledgements}
We gratefully acknowledge
\begin{itemize}
    \item Laurie Rousseau-Nepton and Simon Prunet, for their help with understanding imaging Fourier transform spectrographs
    \item Tyler Robinson, for his help in understanding the various noise terms in the direct imaging error budget
    \item Matthew Bolcar and Stefan Martin, for their help with understanding the LUVOIR and HabEx mission optical designs,
    \item Bertrand Mennesson, for his help in understanding the exozodiacal flux levels and coronagraph performance for space-based exoplanet imaging.
    
\end{itemize}

This work was supported by NASA FINESST grant \#20-ASTRO20-0172/Jingwen Zhang. J.Z. would like to thank Jerry Xuan for helpful discussions. M.B. would also like to particularly thank his undergraduate research advisor, Amber Miller, for introducing him to the fascinating world of Fourier transform spectroscopy.  Part of this work was carried out at the Jet Propulsion Lab, California Institute of Technology, under contract with NASA (80NM00018D0004). We would like to thank the reviewer(s) for the thorough and fair comments, which helped to improve the rigor of many aspects of the paper.

\vspace{5mm}

\software{\textit{AstroPy} \citep{2013A&A...558A..33A},  
          \textit{SciPy} \citep{scipy2020},
          \textit{NumPy} \citep{numpy2020},
          \textit{Pandas} \citep{panda2021}
          }

\appendix
\counterwithin{figure}{section}
\section{Planet light and Astrophysical Background}\label{sec:plab}
We derived analytic expressions for the SNRs of integral field spectrographs and imaging Fourier transform spectrographs in section~\ref{sec:ftsmodel} and section~\ref{sec:ifsmodel}.  These expressions include the planet light, background (local zodiacal light, exzodiacal light, speckle light). In this section, we will present the detailed analysis of each term. The analysis in this section follows the work in \citet{Robinson2016}, \citet{Lacy2019}, and \citet{Stark2019}.  For the IFS cases, we checked these results against their published work, finding excellent agreement in computed count rates. Here, we present the spectra as the function of wavelength, and they can be transformed to the format of wavenumber for iFTS.

\subsection{Planet count rate, $c_p$} 
In the optical and near-infrared, thermal emission from mature  terrestrial and gas giant exoplanets is negligible.  The flux from the planet is due to reflection of stellar light.  We assume the host star has a radius $R_{\star}$ and an effective temperature of $T_{\rm{eff}}$. The stellar specific flux density at the planet's orbit is 

\begin{equation}
    F_{\star,\lambda}(a) = \pi B_{\lambda}(T_{\rm{eff}})\left(\frac{R_{\star}}{a}\right)^{2}
\end{equation}
where $B_{\lambda}(T_{\rm{eff}})$ is the blackbody spectrum, $a$ is the orbital distance of the planet.  More realistic stellar spectra, either from observations or models, can also be used for $ F_{\star,\lambda}(a)$. 

The fraction of stellar light at wavelength $\lambda$ reflected by a planet toward an observer is determined by the planet's radius $R_{p}$, geometric albedo $A_{g}(\lambda)$, and phase function $\Phi(\alpha)$.  The albedo depends on the wavelength-dependent scattering properties of the atmosphere and planetary surface and is of primary scientific interest.  The phase function accounts for the variation of reflectivity over the planet's orbit, with $\alpha$ being the exoplanet-centric angle between the star and the observer  \citep{MadhusudhanandBurrow2012}.  While the phase function may also depend weakly on wavelength \citep{Mayorga2016}, we ignore such effects and use a Lambertian form:

\begin{align}
    \Phi(\alpha)&=\frac{\sin \alpha + (\pi - \alpha)\cos \alpha}{\pi}
\end{align}

\noindent Combining these expressions, we see the specific flux density of a planet at a distance of $d$ from Earth is 
\begin{equation}
\begin{aligned}
    F_{p}(\lambda) &= A_{g}(\lambda)\Phi(\alpha)F_{\star,\lambda}(a)\left(\frac{R_{p}}{d}\right)^{2} \\
                &= \pi B_{\lambda}(T_{\rm{eff}}) A_{g}(\lambda)\Phi(\alpha)\left(\frac{R_{\star}}{a}\right)^{2}\left(\frac{R_{p}}{d}\right)^{2}
\end{aligned}
\end{equation}

We can convert the planet specific flux density in the unit of $W\ m^{-2}\ nm^{-1}$ into photon density in the unit of $e^{-}\ m^{-2}\ nm^{-1}$  by multiplying by $\frac{\lambda}{hc}$. The actual detected photons from the planet also depends on the effective collecting area $A_{\rm{PM}}$ and total system throughput $\tau_{tot}$. Thus, the planetary photons at wavelength $\lambda$ measured by the detector per unit time is  

\begin{equation}
\begin{aligned}
    c_{p}(\lambda) &= A_{\rm{PM}} \tau_{tot}  F_{p}(\lambda) \frac{\lambda}{hc} \\
            &= \pi A_{\rm{PM}} \tau_{tot}
\end{aligned}
\end{equation}

\subsection{Zodiacal and exozodiacal background light}
Zodiacal and exozodiacal light consists of scattered starlight from dust in the solar system and extrasolar systems, respectively, and is a primary source of astrophysical noise at optical and near infrared wavelengths ($0.5-2\ \mu m$). \cite{stark2014} demonstrated that a uniform V-band surface brightness of $M_{\rm{z,V}}=23\ \rm{mag\ arcsec^{\text{-}2}}$ is a reasonable representation of solar system zodiacal light, ignoring the variation with ecliptic latitude and longitude given by \cite{LeRe1980}. The solar system zodiacal light flux density is given by 
\begin{equation}
    F_{z}(\lambda) = \Omega \frac{F_{\rm{\odot,\lambda}}(1\rm{AU)}}{F_{\rm{\odot,V}}(1\rm{AU)}}F_{\rm{0,V}}10^{\text{-}M_{\rm{z,V}}/2.5}
\end{equation}
where $F_{\rm{\odot,\lambda}}$ is the wavelength-dependent specific solar flux density, $F_{\rm{\odot,V}}$ is the solar flux density in the V-band, and $F_{\rm{0,V}}$ is the standard zero-magnitude V-band specific flux density. $\Omega$ is the photometric aperture in units of $\rm{arcsec^{2}}$. Following \cite{Robinson2016}, we use a square aperture area $\Omega = \left(\frac{\lambda}{D}\right)^{2}$ to represent the PSF region of the planet. Thus, the solar system zodiacal light photons at wavelength $\lambda$ received by the detector per unit time is 
\begin{equation}
\begin{aligned}
    c_{z}(\lambda) &=  A_{\rm{PM}} \tau_{tot}F_{z}(\lambda)\frac{\lambda}{hc} \\
    &=  A_{\rm{PM}} \tau_{tot}\Omega \frac{F_{\rm{\odot,\lambda}}(1\rm{AU)}}{F_{\rm{\odot,V}}(1\rm{AU)}}F_{\rm{0,V}}10^{\text{-}M_{\rm{z,V}}/2.5}\frac{\lambda}{hc}
\end{aligned}
\end{equation}

Following \cite{stark2014}, a ``zodi'' is defined as the surface brightness of an exozodiacal disk at $1\ \rm{AU}$ from a solar twin. The surface brightness in V band $M_{\rm{ex,V}}$ is 22 $\rm{mag\ arcsec^{\text{-}2}}$, which is roughly twice the local zodi level, since the observer receives reflected light of dust from both above and below the mid-plane of the system. Therefore, the exozodiacal light flux density from a system with $N_{ex}$ zodis of dust  around a star with effective temperature $T_{\rm{eff}}$ at the planet orbit is given by 
\begin{equation}
\begin{aligned}
    F_{ez}(\lambda)&=N_{ex}\Omega \frac{F_{\star,\lambda}(a)}{F_{\star,V}(1\rm{AU})} \frac{F_{\star,V}(1\rm{AU})}{F_{\odot,V}(1\rm{AU})}\\
    &\times F_{\rm{0,V}}10^{\text{-}M_{\rm{ez,V}}/2.5} \\
                   &= N_{ex}\Omega \frac{\pi B_{\lambda}(T_{\rm{eff}})}{F_{\odot,V}(1\rm{AU})}\left(\frac{R_{\star}}{a}\right)^{2}\\ &\times F_{\rm{0,V}}10^{\text{-}M_{\rm{ez,V}}/2.5}
\end{aligned}
\end{equation}

\noindent where $a$ is the planet's orbital distance. The term $\frac{F_{\star,V}}{F_{\odot,V}}$ accounts for the exozodical light scaling with the brightness of the host star.  Notice the exozodical light decreases with the increasing orbital distance as $1/a^{2}$.

The exozodiacal light photons received at wavelength $\lambda$ by the detector per unit time is
\begin{equation}
\begin{aligned}
    c_{ez}(\lambda)&= A_{\rm{PM}} \tau_{tot}F_{ez}(\lambda)\frac{\lambda}{hc} \\
                   &= A_{\rm{PM}} \tau_{tot}N_{ex}\Omega \frac{\pi B_{\lambda}(T_{\rm{eff}})}{F_{\odot,V}(1\rm{AU})}\left(\frac{R_{\star}}{a}\right)^{2}\\ &\times F_{\rm{0,V}}10^{\text{-}M_{\rm{ez,V}}/2.5}\frac{\lambda}{hc} 
\end{aligned}
\end{equation}
The above expressions reveal that the solar system zodiacal and exozodiacal light signals are independent of the telescope diameter. But the ratio of the planetary signal to the solar system zodiacal and exozodiacal light signals scale with telescope size as $c_{p}/(c_{z}+c_{ez})\propto D^{2}$. Thus, a larger telescope helps to distinguish the planet light from these sources of background noise. 
\begin{figure}
    \centering
    \includegraphics[width=0.5\linewidth]{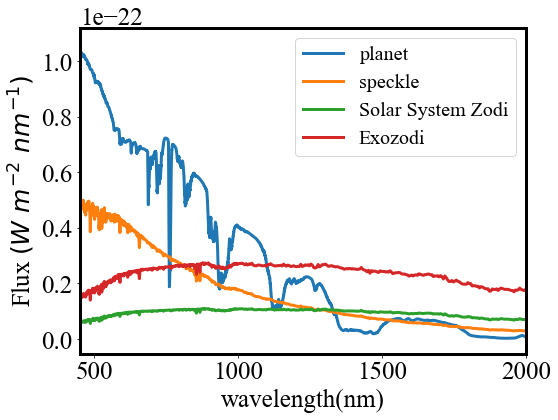}
    \caption{ The specific flux density of  planet, zodiacal light, exozodiacal light, and speckle light at the pupil of a telescope for an Earth twin orbiting a Sun-like star 10 pc away.  Here, the zodiacal and exozodiacal light are 23 and 22 $\rm{mag\ arcsec^{-2}}$, respectively. The simulated telescope has a 6 m primary mirror and achieves a contrast of $10^{-10}$.}
    \label{fig:figure3}
\end{figure}
\subsection{Speckle Light}
Another astrophysical noise source arises from the interaction of stellar light with unsensed or uncorrected optical aberrations in the telescope and coronagraph, causing a relatively bright ``speckle'' halo surrounding the image of the star in the focal plane.  The speckle flux density is the product of stellar specific flux density and the coronagraph raw contrast $C_{\rm{raw}}$:
\begin{equation}
\begin{aligned}
    F_{sp}(\lambda)&=C_{\rm{raw}}F_{\star,\lambda}(d)\\
                    &=C_{\rm{raw}}\pi B_{\lambda}(T_{\rm{eff}})\left(\frac{R_{\star}}{d}\right)^{2}
\end{aligned}
\end{equation}
Thus, speckle light photons at wavelength $\lambda$ received by the detector per unit time is
\begin{equation}
\begin{aligned}
    c_{sp}(\lambda)&= A_{\rm{PM}} \tau_{tot}F_{\rm{sp}}(\lambda)\frac{\lambda}{hc}\\
                    &=A_{\rm{PM}} \tau_{tot}C_{\rm{raw}}\pi B_{\lambda}(T_{\rm{eff}})\left(\frac{R_{\star}}{d}\right)^{2}\frac{\lambda}{hc}
\end{aligned}
\end{equation}

Figure~\ref{fig:figure3} shows the specific flux density of each term at the telescope pupil from 450-2000 nm for a Earth twin orbiting a Sun-like star at 1 AU from 10 pc away. 

\section{System Throughput}\label{sec:ST}

The system throughput $\tau_{tot}$ is the product of optical throughput $\tau_{\rm{optical}}$, coronagraph core throughput  $\tau_{\rm{core}}$ and detector quantum efficiency $\tau_{\rm{QE}}$.  Here, we briefly describe these parameters.

\subsection{Optical throughput}

The optical throughput of the instrument depends on the various reflectivities of the surfaces, transmissive and coupling losses (such as from lenslets), etc.  Generally, optical throughput is a weak function of separation between the star and planet, but can be somewhat dependent on wavelength. In our simulations, we use the optical throughput in optical ($24\%$) and near-IR ($32\%$) from the LUVOIR report \citep{LUVOIRTeam}. We use the same the optical throughput of the IFS and iFTS in the simulations.

\begin{figure}
    \centering
    \includegraphics[width=0.9\linewidth]{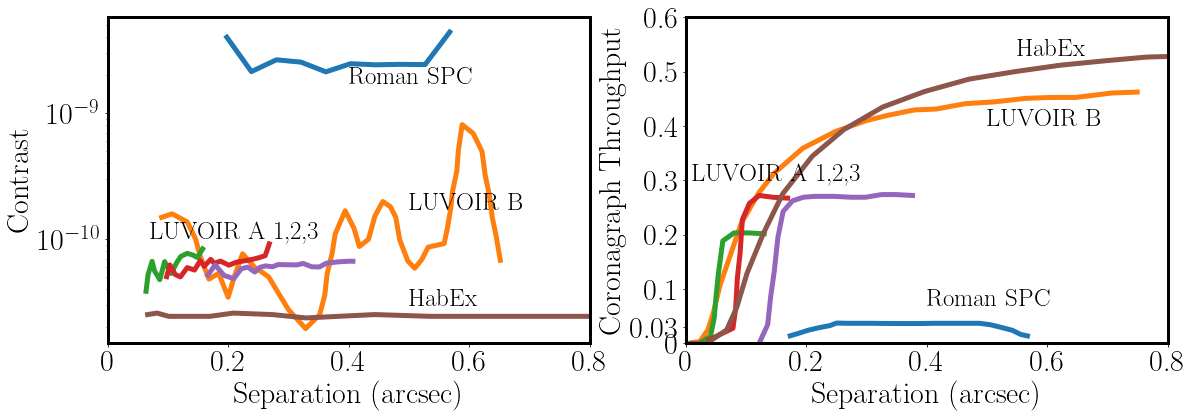}
    \caption{Azimuthal mean raw contrast and coronagraph core throughput $\tau_{core}$ of Roman-SPC, LUVOIR A-APLC, LUVOIR B-DMVC and HabEx as a function of separation. LUVOIR A 1,2,3 correspond to three adopoted masks desinged for LUVOIR A. Sources: Roman-SPC \citep{Lacy2019}, LUVOIR \citep{Stark2019}, Habex \citep{Ruane2018} }
    \label{fig:figure4}
\end{figure}
\subsection{Coronagraph core throughput}
The intrinsic faintness of the signals involved in exoplanet imaging leads to the conclusion that it is not realistic to expect to detect photons from the planet outside the central core of the planet's PSF.  This leads to the concept of ``core throughput'', which captures both the diffraction of the light outside the center of the planet PSF due to the coronagraph as well as any throughput losses from light being blocked by the coronagraph optics (for example, an opaque hard-edge Lyot coronagraph will heavily attenuate point sources located behind the focal plane mask).  Core throughput is a weak function of wavelength, but strongly dependent on the separation between the planet and the star. Figure~\ref{fig:figure4} shows the contrast and core throughput of different coronagraphs as a function of separation.

\subsection{Detector quantum efficiency}\label{sec:Dqe}
Detector quantum efficiency, or QE, is the final component of throughput, which depends on the detector substrate and coatings.  QE can be a strong function of wavelength. In the optical, silicon detectors are typically used, which have excellent QE in visible wavelengths but become increasingly transparent at infrared wavelengths.  To capture this effect, we used the quantum efficiency for the visible detector from \cite{Lacy2019}. In the infrared, HgCdTe has become the substrate of choice.  While the QE wavelength cutoffs may be adjusted by changing the relative abundances of the elements, the QE at near-infrared wavelengths is well approximated by a constant 80\% from 1-2 $\mu$m, which is the functional form we adopt.
\begin{equation}
    \tau_{\rm{QE}} =
    \begin{cases}
      0.9, & \lambda \leq 0.7 \mu m\\
      0.9(1-\frac{\lambda-0.7}{0.3}), & 0.7<\lambda \leq 1 \mu m\\
      0.8, & 1<\lambda \leq 2 \mu m
    \end{cases} 
\end{equation}

\section{Practical observation of spectra using an iFTS}\label{sec:POS}

\subsection{Aliasing}\label{subsec:alis}
A truncated, discrete interferogram inevitably leads to some spectral distortions compared to an infinitely long, continuous signal. One distortion due to the discrete sampling is aliasing. In section~\ref{sec:FTSconcept} and section~\ref{sec:ftsmodel}, we have discussed that the output spectrum from the Fourier transform to the interferogram is a symmetric mirror spectrum along the zero point. However, we can see Eqn.~\ref{eqn:ftsout} is not only valid for indices $i$ from $-N$ to $N$, but also for all other integers. If we replace $\nu_{i}$ with $\nu_{i}+{n}/{\Delta \delta}$, where $n$ can be any non-zero integer, we see that

\begin{equation}
\begin{aligned}
    O_{FTS}\left(\nu_{i}+\frac{n}{\Delta \delta}\right) &= \Delta \delta \sum_{-N}^{N} F_{\Delta}(\delta_{j})\cos\left(-2\pi\left(\nu_{i}+\frac{n}{\Delta \delta}\right) \Delta \delta\right)\\
                                    &=\Delta \delta \sum_{-N}^{N} F_{\Delta}(\delta_{j})\cos(-2\pi\nu_{i} \Delta \delta)\\
                                    &=O_{FTS}(\nu_{i})
\end{aligned}
\end{equation}

\begin{figure}
    \centering
    \includegraphics[width=0.6\linewidth]{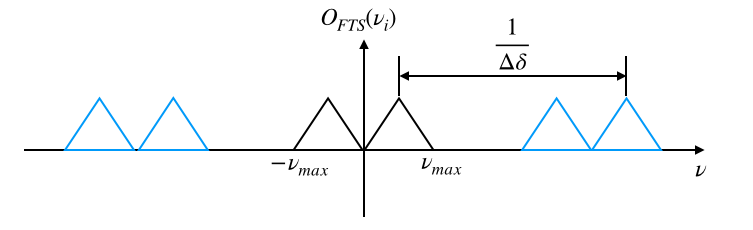}
    \caption{The output spectra from Fourier transform repeat at an interval of $\frac{1}{\Delta \delta}$. }
    \label{fig:figurea4}
\end{figure}
\noindent which means that the mirror-symmetrical spectrum computed from an interferogram sampled with the interval $\Delta \delta$ is periodic with the period $\frac{1}{\Delta \delta}$ as shown in Figure~\ref{fig:figurea4}. The replication of the original spectrum and its mirror images on the wavenumber axis is called aliasing. It is clear that the original spectrum can only be obtained if it does not overlap with the replications. Thus, the critical sampling interval of an interferogram is 
\begin{equation}
    (\Delta \delta)_{\rm{Nyquist}} = \frac{1}{2\nu _{\rm{max}}}
\end{equation}
where $\nu _{\rm{max}}$ is the maximum wave number (ie, smallest wavelength) of the spectral band. For example, for a spectrum with a minimum wavelength of 400 nm, the optical path difference would need to be 200 nm or less, ie, a mirror step of 100 nm.  If the interferogram is sampled at exactly the critical sampling interval, the spectrum and its mirror image replicate with the period of $2\nu_{max}$. If the sampling interval is smaller than this critical interval $\Delta \delta  < \frac{1}{2\nu _{\rm{max}}}$, then the spectral orders of the sampled signal do not overlap and the original spectrum can be faithfully recreated. In contrast, if the sampling interval $\Delta \delta > \frac{1}{2\nu _{\rm{max}}}$, the period of the replications is smaller than $2\nu_{\rm{max}}$. Then the portion of the spectrum with $|\nu|> \nu_{\rm{max}}$ is aliased into the
basic period $[-\nu_{\rm{max}}, \nu_{\rm{max}}]$ and thus overlaps with the spectral information originally located in this interval. In this case, we obtain a distorted spectrum. 

In summary, the interferogram sampling interval needs to be smaller than the Nyquist interval  $\frac{1}{2\nu_{\rm{max}}} $ to avoid the aliasing.

\subsection{Spectral resolving power}\label{sec:resolution}
The finite mirror scan distance in the range $(-L, L)$, where $L=N\Delta \delta $, leads to a truncation of the interferogram.  This results in the convolution of the true spectrum with an instrument profile, similarly to how a dispersive spectrograph convolves the true spectrum with a line-spread function that depends on the resolving power. This can be derived as follows: the truncated interferogram $I_{L}(\delta)$ can be written as the product of the ``infinite'' interferogram $I_{\Delta}(\delta)$ and a boxcar function:
\begin{equation}
     I_{L}(\delta)=\Pi(\delta)I_{\Delta}(\delta) 
\end{equation}
where $\Pi(\delta)$ is the boxcar function: 
\begin{equation}
    \Pi(\delta)=
        \begin{cases}
          1, & |x|\leq L,\\
          0, & |x| > L
        \end{cases} 
\end{equation}

\begin{figure*}
    \centering
    \includegraphics[width=0.6\linewidth]{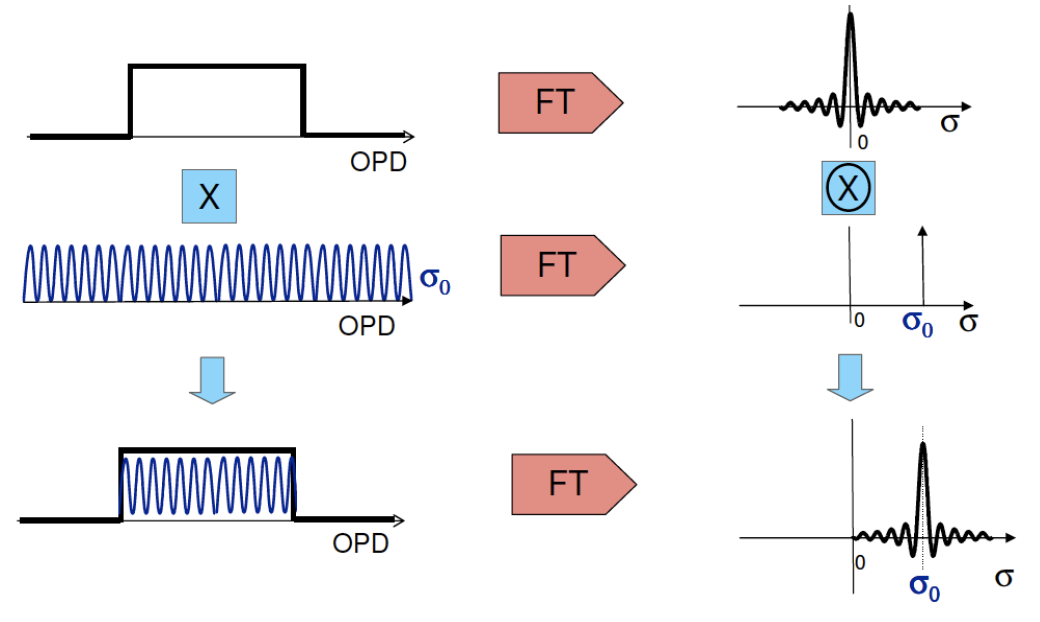}
    \caption{The truncation effect of the interferogram. The measured spectrum is the convolution of the original spectrum with the instrument profile, which is the Fourier transform of a boxcar function.Figure is adapted from \cite{Drissen2001}}
    \label{fig:figurea2}
\end{figure*}
Based on the properties of the Fourier transform, we can see that the spectrum computed from the truncated interferogram $B_{L}(\nu)$ is the convolution of true spectrum $B(\nu)$ and the inverse Fourier transform of the boxcar function\begin{equation}
    \begin{split}
    B_{L}(\nu) & = \mathcal{F}^{\text{-}1}\{ \Pi(\delta)I_{\Delta}(\delta)\} =\mathcal{F}^{\text{-}1}\{\Pi(\delta)\}\ast \mathcal{F}^{\text{-}1}\{I_{\Delta}(\delta)\} \\
    & = \mathcal{F}^{\text{-}1}\{\Pi(\delta)\}\ast B(\nu)
    \end{split}
\end{equation}

The inverse Fourier transform of the boxcar function is a sinc function, $2L \rm{sinc}(2\pi\nu L)$. Thus, the effect of truncation of the interferogram is that the true spectrum is convolved by a sinc function, also called the instrumental line shape. For example, if we measured a monochromatic beam, the computed spectrum using truncated interferogram would be a sinc functions instead of a ideal delta function (see Figure~\ref{fig:figurea2}). 

The full width at half maximum (FWHM) of the instrumental line shape is $\Delta \nu=1.207/(2L)$. Thus, the resolving power can be given by 
\begin{equation}\label{res_equation}
    R=\frac{\lambda}{\Delta \lambda}=\frac{\lambda}{\lambda^{2}\Delta \nu}=\frac{2L}{1.207\lambda} = \frac{1.65L}{\lambda}
\end{equation}
 Thus, one can modify the resolving power of an iFTS by changing the maximum scan length.\footnote{There is a small subtlety here, in that an iFTS delivers a fixed resolving power in \textit{wavenumber}, which is slightly different than saying the quantity $\lambda/\Delta \lambda$ is fixed. Comparing the performance of an iFTS and IFS necessarily means matching the resolving power at the center of the filter.  The variation in resolving power along the filter is usually about 5-10\%.}

\begin{figure*}[b]
    \centering
    \includegraphics[width=0.8\linewidth]{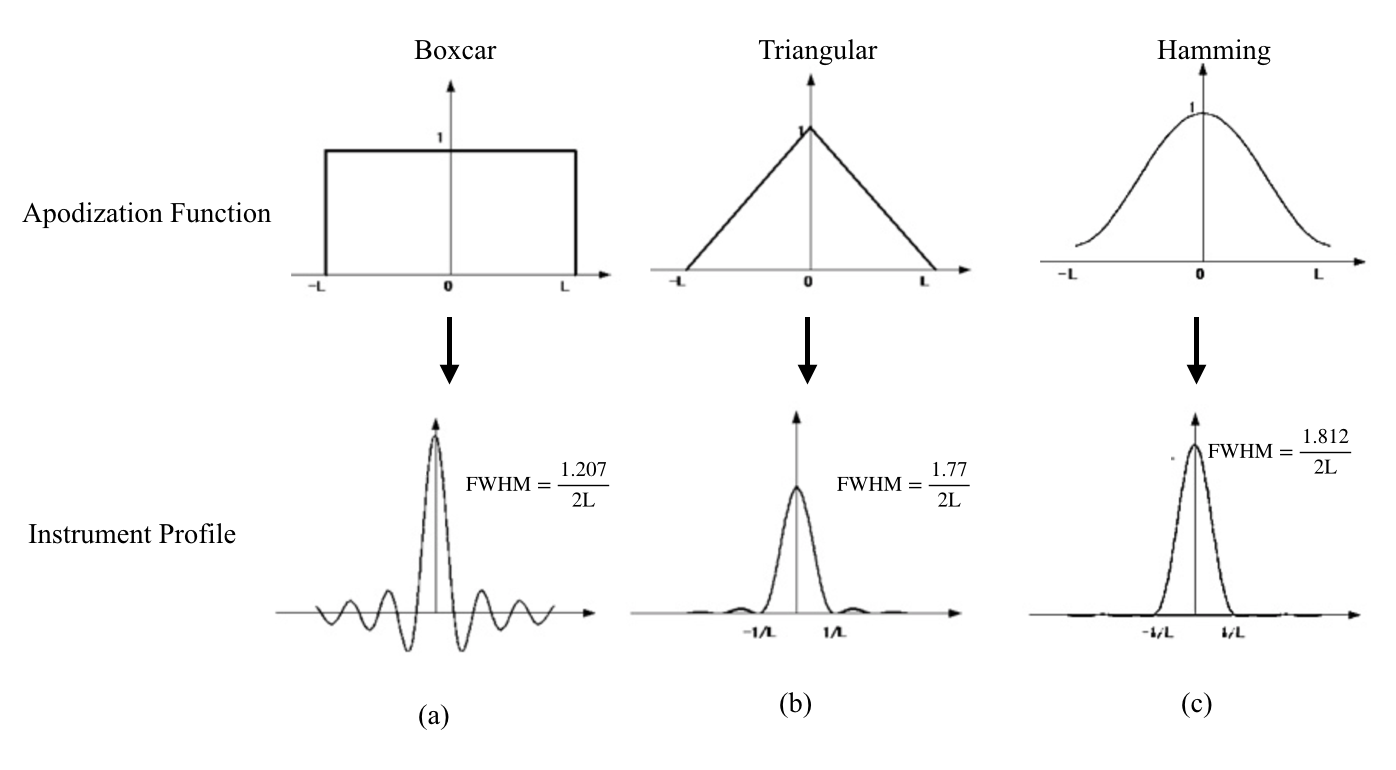}
    \caption{Boxcar, triangular and hamming apodization functions and their instrument profiles.  }
    \label{fig:figurea3}
\end{figure*}

\subsection{Apodization}\label{sec:ap}
 For a basic boxcar scan pattern (equal time per step) the corresponding instrument profile is a sinc function shown in the Figure~\ref{fig:figurea3}(a). We can see besides a main lobe centered at 0, there are numerous additional peaks called side lobes which cause a leakage of the spectral intensity. For the sinc function, the amplitude of the largest side lobe is $22\%$ of that of the main lobe. Because side lobes do not contribute to the measured spectral information but represent the instrument pattern due to the truncation of signal, it is desirable to reduce their amplitude. 
 
 The method to attenuate the side lobes is called apodization. The easiest way to do this is by modifying the time spent per step in mirror position. Figure~\ref{fig:figurea3} (b) and (c) present a triangular apodization function and a Hamming apodization function, respectively. Compared to the boxcar function, the instrument profiles of these functions have a wider main lobe for the same scan length, leading to higher signal-to-noise ratio and lower resolving power. The inverse Fourier transform of the triangular function is $L sinc^{2}(\pi \nu L)$, which has a FWHM of $\frac{1.77}{2L}$, ie, about 30\% wider. Another option is the Hamming function, $0.54 - 0.45\cos(2\pi x/L)$ with a FWHM of $\frac{1.812}{2L}$. There has been significant research into optimal apodization functions \citep{naylor2007apodizing}, and the choice can  depend on the science goals, and on the strength and number of the lines.


\clearpage

\bibliography{sample63}{}
\bibliographystyle{aasjournal}


\end{CJK*}
\end{document}